\documentclass[prd,aps,showkeys,floats,onecolumn]{revtex4}
\usepackage{amsmath}
\usepackage{amssymb}

\begin{document}

\title{A review about Invariance Induced Gravity: Gravity and Spin from Local Conformal-Affine Symmetry}

\author{ S. Capozziello and M. De Laurentis}
\affiliation{ Dipartimento di Scienze Fisiche, Universit\`a di
Napoli "Federico II" and INFN Sez. di Napoli, Compl. Univ. Monte
S. Angelo, Ed.N, Via Cinthia, I-80126 Napoli, Italy}

\begin{abstract}
In this review paper, we discuss how gravity and spin can be
obtained as the realization of the local Conformal-Affine group of
symmetry transformations. In particular, we show how gravitation
is a gauge theory which can be obtained starting from some local
invariance as the Poincar\'{e} local symmetry. We review previous
results where the inhomogeneous connection coefficients,
transforming under the Lorentz group, give rise to gravitational
gauge potentials which can be used to define covariant derivatives
accommodating minimal couplings of matter, gauge fields (and then
spin connections). After we show, in a self-contained approach,
how the tetrads and the Lorentz group can be used to induce the
spacetime metric and then the {\it Invariance Induced Gravity} can
be directly obtained both in holonomic and anholonomic pictures.
Besides, we show how tensor valued connection forms act as
auxiliary dynamical fields associated with the dilation, special
conformal and deformation (shear) degrees of freedom, inherent to
the bundle manifold. As a result, this allows to determine the
bundle curvature of the theory and then to construct boundary
topological invariants which give rise to a prototype (source
free) gravitational Lagrangian. Finally, the Bianchi identities,
the covariant field equations and the gauge currents are  obtained
determining completely the dynamics.
\end{abstract}


\maketitle

\section{Introduction}

General Relativity and Quantum Mechanics  are  two fundamental
theories of modern physics and the  Standard Model of particles is
currently the most successful relativistic quantum field theory.
It is a non-Abelian gauge
theory (Yang-Mills theory) associated with the internal symmetry group $%
SU(3)\times SU(2)\times U(1)$, in which the $SU(3)$ color symmetry for the
strong force in quantum chromodynamics is treated as exact whereas the $%
SU(2)\times U(1)$ symmetry, responsible for generating the
electro-weak gauge fields, is spontaneously broken. So far as we
know, there are four fundamental forces in Nature; namely,
electromagnetic force, weak force, strong force and gravitational
force. The Standard Model covers the first three, but not the
gravitational interaction.

Here we intend to give a short summary of the various attempts to
put together gravitation an the other interactions in view of a
self-contained unified theory.

In General Relativity, the geometrized gravitational field is
described by the metric tensor $g_{\mu \nu }$ of pseudo-Riemannian
spacetime, and the field equations that the metric tensor
satisfies are nonlinear. This nonlinearity is indeed a source of
difficulty in quantization of General Relativity. Since the
successful Standard Model of particle physics is a gauge theory in
which all the fields mediating the interactions are represented by
gauge potentials, a question arises to understand why the fields
mediating the gravitational interaction are different from those
of other fundamental forces. It is reasonable to expect that there
may be a gauge theory in which the gravitational fields stand on
the same footing as those of other fields. This expectation has
prompted a re-examination of General Relativity from the gauge
theoretical point of view.

While the gauge groups involved in the Standard Model are all
internal symmetry groups (e.g. spin is an internal symmetry), the
gauge groups in General Relativity must be associated to external
spacetime symmetries. Therefore, the gauge theory of gravity would
not be a usual Yang-Mills theory. It must be one in which gauge
objects are not only the gauge potentials but also tetrads that
relate the symmetry group to the external spacetime. For this
reason, we have to consider a more complex nonlinear gauge theory.
In General Relativity, Einstein took the spacetime metric as the
basic variable representing gravity, whereas Ashtekar employed the
tetrad fields and the connection forms as the fundamental
variables. We also consider the tetrads and the connection forms
as the fundamental fields.

R. Utiyama  was the first to suggest that gravitation may be
viewed as a gauge theory \cite{Utiyama} in analogy to the
Yang-Mills theory \cite{YangMills}. He identified the gauge
potential, due to the Lorentz group, with the symmetric connection
of Riemann geometry, and constructed Einstein's General Relativity
as a gauge theory of the Lorentz group $SO(3$, $1)$ with the help
of tetrad fields introduced in an \textit{ad hoc} manner. Although
the tetrads were necessary components of the theory to relate the
Lorentz group, adopted as an internal gauge group to the external
spacetime, they were not introduced as gauge fields. After, Kibble
\cite{Kibble}
constructed a gauge theory based on the Poincar\'{e} group $P(3$, $%
1)=T(3$, $1)\rtimes SO(3$, $1)$ ($\rtimes $ represents the
semi-direct product) which resulted in the Einstein-Cartan theory
characterized by curvature and torsion. The translation group
$T(3$, $1)$ is considered responsible for generating the tetrads
as gauge fields. Cartan \cite{Cartan} generalized the Riemann
geometry to include torsion in addition to curvature. The torsion
(tensor) arises from an asymmetric connection. Sciama
\cite{Sciama}, and others (R. Fikelstein \cite{Finkelstein}, Hehl
\cite{Hehl1, Hehl2}) pointed out that intrinsic spin may be the
source of torsion of the underlying spacetime manifold.

Since the form and role of the tetrad fields are very different
from those of gauge potentials, it has been thought that even
Kibble's attempt is not satisfactory as a full gauge theory. There
have been a number of gauge theories of gravitation based on
several Lie groups \cite{Hehl1, Hehl2, Mansouri1, Mansouri2,
Chang, Grignani, MAG}. It was argued that a gauge theory of
gravitation, corresponding to General Relativity, can be
constructed with the translation group alone in the so-called
teleparallel scheme. Inomata \textit{et al.} \cite{Inomata}
proposed that Kibble's gauge theory could be obtained in a way
closer to the Yang-Mills approach by considering the de Sitter
group $SO(4$, $1)$ which is reducible to the Poincar\'{e} group by
group-contraction. Unlike the Poincar\'{e} group, the de Sitter
group is homogeneous and the associated gauge fields are all of
gauge potential type. By the Wigner-In\"{o}nu group contraction
procedure, one of the five vector potentials reduces to the
tetrad.

It is common to use the fiber-bundle formulation by which gauge
theories can be constructed on the basis of any Lie group. A work
by Hehl \textit{et al.} \cite{MAG} on the so-called Metric-Affine
Gravity  adopted,
as a gauge group, the affine group $A(4$, $\mathbf{%
%TCIMACRO{\U{211d} }%
%BeginExpansion
\mathbb{R}
%EndExpansion
})=T(4)\rtimes GL(4$, $\mathbf{%
%TCIMACRO{\U{211d} }%
%BeginExpansion
\mathbb{R}
%EndExpansion
})$ which was realized linearly. The tetrad was identified with
the nonlinearly realized translational part of the affine
connection on the tangent bundle. In metric-affine gravity, the
Lagrangian is quadratic in both curvature and torsion in contrast
to the Einstein-Hilbert Lagrangian of General Relativity which is
linear in the scalar curvature. The theory has the Einstein limit
on one hand and leads to the Newtonian inverse distance potential
plus the linear confinement potential in the weak field
approximation on the other. This approach has been recently
developed also for more general theories as $f(R)$-gravity (see
\cite{stefano1,stefano2} and also \cite{francaviglia}). As we have
seen above, there are many attempts to formulate gravitation as a
gauge theory. Currently no theory has been uniquely accepted as
the gauge theory of gravitation.

The nonlinear approach to group realizations was originally
introduced by  Coleman,  Wess and  Zumino \cite{CCWZ1, CCWZ2} in
the context of internal symmetry groups. It was later extended to
the case of spacetime symmetries by Isham, Salam, and Strathdee
\cite{Isham, Salam}
considering the nonlinear action of $GL(4$, $\mathbf{%
%TCIMACRO{\U{211d} }%
%BeginExpansion
\mathbb{R}
%EndExpansion
})$ mod the Lorentz subgroup. In 1974, Borisov, Ivanov and Ogievetsky \cite%
{BorisovOgievetskii, IvanovOgievetskii} considered the
simultaneous nonlinear realization  of the affine and conformal
groups. They showed that General Relativity can be viewed as a
consequence of spontaneous breakdown of the affine symmetry in
much the same manner that chiral dynamics in quantum
chromodynamics is a result of spontaneous breakdown of chiral
symmetry. In their model, gravitons are considered as Goldstone
bosons associated with the affine symmetry breaking. In 1978,
Chang and
Mansouri \cite{ChangMansouri} used the nonlinear realization scheme employing $GL(4$, $\mathbf{%
%TCIMACRO{\U{211d} }%
%BeginExpansion
\mathbb{R}
%EndExpansion
})$ as the principal group. In 1980, Stelle and West
\cite{StelleWest} investigated the nonlinear realization induced
by the spontaneous breakdown of $SO(3$, $2)$. In 1982 Ivanov and
Niederle considered nonlinear gauge theories of the
Poincar\'{e}, de Sitter, conformal and special conformal groups \cite%
{IvanovNiederle1, IvanovNiederle2}. In 1983, Ivanenko and Sardanashvily \cite%
{IvanenkoSardanashvily} considered gravity to be a spontaneously broken $%
GL(4 $, $\mathbf{%
%TCIMACRO{\U{211d} }%
%BeginExpansion
\mathbb{R}
%EndExpansion
})$ gauge theory. The tetrads fields arise in their formulation as
a result of the reduction of the structure group of the tangent
bundle from the general linear  Lorentz group. In 1987, Lord and
Goswami \cite{Lord1, Lord2} developed the nonlinear realization in
the fiber bundle formalism based on the bundle structure $G\left(
G/H\text{, }H\right) $ as suggested by Ne'eman and Regge
\cite{NeemanRegge}. In this approach the quotient space $G/H$ is
identified with physical spacetime. Most recently, in a series of
papers, A. Lopez-Pinto, J. Julve, A. Tiemblo, R. Tresguerres and
E. Mielke discussed nonlinear gauge theories of gravity on the
basis of the Poincar\'{e}, affine and conformal groups
\cite{Julve, Lopez-Pinto, TresguerresMielke, Tresguerres,
TiembloTresguerres1, TiembloTresguerres2}.

Now, following the prescriptions of General Relativity, the
physical spacetime is assumed to be a four-dimensional
differential manifold. In Special Relativity, this manifold is the
Minkwoski flat-spacetime $M_{4}$ while, in General Relativity, the
underlying spacetime is assumed to be curved in order to describe
the effects of gravitation.

As we said, Utiyama \cite{Utiyama}  proposed that General
Relativity can be seen as a gauge theory based on the local
Lorentz group  in the same way that the Yang-Mills gauge theory
 \cite{YangMills} is developed on the basis of the internal
iso-spin gauge group. In this formulation the Riemannian
connection is the gravitational counterpart of the Yang-Mills
gauge fields. While $SU(2)$, in the Yang-Mills theory, is an
internal symmetry group, the Lorentz symmetry represents the local
nature of spacetime rather than internal degrees of freedom. The
Einstein Equivalence Principle, asserted for General Relativity,
requires that the local spacetime structure can be identified with
the Minkowski spacetime possessing Lorentz symmetry.

In order to relate local Lorentz symmetry to the external
spacetime, we need to solder the local space to the external
space. The soldering tools can be the tetrad fields. Utiyama
regarded the tetrads as objects given \textit{a priori} while they
can be dynamically generated \cite{unification} and the spacetime
has  necessarily to be endowed with torsion in order to
accommodate spinor fields. In other words, the gravitational
interaction of spinning particles requires the modification of the
Riemann spacetime of General Relativity to be a (non-Riemannian)
curved spacetime with torsion. Although Sciama used the tetrad
formalism for his gauge-like handling of gravitation, his theory
fell shortcomings in treating tetrad fields as gauge fields.
Following the Kibble approach \cite{Kibble}, it can be
demonstrated how gravitation can be formulated starting from a
pure gauge viewpoint.

After this short summary of thirty years long attempts to put
General Relativity on the same footing of non-Abelian gauge
theories,  the aim of this paper is to show, in details, how a
theory of gravitation is a gauge theory which can be obtained
starting from some local invariance, e.g. the local Poincar\'{e}
symmetry (see \cite{felix} and references therein). This dynamical
structure give rise to  a gauge theory of gravity, based on a
nonlinear realization of the local conformal-affine group of
symmetry transformations and on conservation principles
\cite{symplectism}. In particular, we want to show how invariance
properties and conservation laws induce the gravitational field
and internal (spin) fields  generalizing results in
\cite{felix,deformazioni}.

Specifically, in the review part of the paper, we are going to
consider the general problem of how gravity, as a gauge theory,
could be achieved by the nonlinear realization of the
conformal-affine group. We give all the mathematical tools for
this realization discussing in details the bundle approach to the
gauge theories and  investigating also how internal symmetries as
spin could be achieved under the same standard. The result is the
Invariance Induced Gravity which can be seen also as a deformation
of local Poincar\'{e} Gauge Invariance. In this sense, we are
going to complete the discussion in \cite{felix}.

The layout of the paper is the following. In Sect.II,  the
standard bundle approach to the gauge theories is presented.
Sect.III is devoted to the discussion of the bundle structure of
gravitation, while the conformal-affine Lie algebra is introduced
in Sect.IV. The group action and the bundle morphisms are
discussed in Sect.V. In Sect.VI, a generalized gauge
transformation law enabling the gauging of external spacetime
groups is introduced. Demanding that tetrads be obtained as gauge
fields requires the implementation of a nonlinear realization of
the conformal-affine group. Such a nonlinear realization is
carried out over the quotient space $CA(3$, $1)$/$SO(3$, $1)$. The
covariant coset field transformations are discussed in Sect.VII In
Sect.VIII, the general form of the gauge connections of the theory
along with their transformation laws are obtained starting from
their decomposition. Furthermore, we present the explicit
structure of the conformal-affine connections. The nonlinear
translational connection coefficient (transforming as a
$4$-covector under the Lorentz group) is identified as a coframe
field. After the detailed description of mathematical tools, in
Sect.IX, we start with the physical realization of the approach.
In particular, the tetrad components of the coframe are used,
together with the Lorentz group metric, to induce the effective
spacetime metric (in this sense, we can speak about an {\it
Invariance Induced Gravity}). As follows in Sect.X, the bundle
curvature of the theory, together with the variations of its
corresponding field strength components, are determined through
the Cartan structure equations. The Bianchi identities are
obtained in Sect.XI. In Sect.XII, surface ($3D$) and bulk ($4D$)
topological invariants are constructed. The bulk terms (obtained
via exterior derivation of the surface terms) provide a means of
"deriving" a prototype (source free) gravitational action (after
appropriately distributing Lie star operators). The covariant
field equations and gauge currents are finally obtained.
Conclusions are presented in Sect. XIII.

\section{The bundle approach to the gauge theories}

Let us start by briefly reviewing the standard bundle approach to
gauge theories. The bundle formalism, together with the
conformal-affine group, will give us the mathematical tools to
achieve gravity as an interaction induced from invariance
properties. Besides, internal degrees of freedom of
conformal-affine algebra will be related to the spin fields.

First of all, one has to verify that the usual gauge potential
$\Omega$ is the pullback of connection 1-form $\omega $ by local
sections of the bundle. After, the transformation laws of the
$\omega $ and $\Omega $ under the action of the structure group
$G$ can be deduced.

Modern formulations of gauge field theories are expressible geometrically in
the language of principal fiber bundles. A fiber bundle is a structure $%
\left\langle \mathbb{P}\text{, }M\text{, }\pi \text{; }\mathbb{F}%
\right\rangle $ where $\mathbb{P}$ (the total bundle space) and $M$ (the
base space) are smooth manifolds, $\mathbb{F}$ is the fiber space and the
surjection $\pi $\ (a canonical projection) is a smooth map of $\mathbb{P}$
onto $M$,%
\begin{equation}
\pi :\mathbb{P}\rightarrow M\text{.}
\end{equation}%
The inverse image $\pi ^{-1}$ is diffeomorphic to $\mathbb{F}$%
\begin{equation}
\pi ^{-1}\left( x\right) \equiv \mathbb{F}_{x}\approx \mathbb{F}\text{,}
\end{equation}%
and is called the fiber at $x\in M$. The partitioning $\bigcup\nolimits_{x}%
\pi ^{-1}\left( x\right) =\mathbb{P}$ is referred to as the fibration. Note
that a smooth map is one whose coordinatization is $C^{\infty }$
differentiable; a smooth manifold is a space that can be covered with
coordinate patches in such a manner that a change from one patch to any
overlapping patch is smooth, see A. S. Schwarz \cite{Schwarz}. Fiber bundles
that admit decomposition as a direct product, locally looking like $\mathbb{%
P\approx }M\times \mathbb{F}$, is called trivial. Given a set of open
coverings $\left\{ \mathcal{U}_{i}\right\} $ of $M$ with $x\in \left\{
\mathcal{U}_{i}\right\} \subset M$ satisfying $\bigcup\nolimits_{\alpha }%
\mathcal{U}_{\alpha }=M$, the diffeomorphism map is given by%
\begin{equation}
\chi _{i}:\mathcal{U}_{i}\times _{M}G\rightarrow \pi ^{-1}(\mathcal{U}%
_{i})\in \mathbb{P}\text{,}  \label{diff}
\end{equation}%
($\times _{M}$ represents the fiber product of elements defined over space $%
M $) such that $\pi \left( \chi _{i}\left( x\text{, }g\right) \right) =x$
and $\chi _{i}\left( x\text{, }g\right) =\chi _{i}\left( x\text{, }\left(
id\right) _{G}\right) g=\chi _{i}\left( x\right) g\ \forall x\in \left\{
\mathcal{U}_{i}\right\} $ and $g\in G$. Here, $\left( id\right) _{G}$
represents the identity element of group $G$. In order to obtain the global
bundle structure, the local charts $\chi _{i}$ must be glued together
continuously. Consider two patches $\mathcal{U}_{n}$ and $\mathcal{U}_{m}$
with a non-empty intersection $\mathcal{U}_{n}\cap \mathcal{U}_{m}\neq
\emptyset $. Let $\rho _{nm}$ be the restriction of $\chi _{n}^{-1}$ to $\pi
^{-1}(\mathcal{U}_{n}\cap \mathcal{U}_{m})$ defined by $\rho _{nm}:\pi ^{-1}(%
\mathcal{U}_{n}\cap \mathcal{U}_{m})\rightarrow (\mathcal{U}_{n}\cap
\mathcal{U}_{m})\times _{M}G_{n}$. Similarly let $\rho _{mn}:\pi ^{-1}(%
\mathcal{U}_{m}\cap \mathcal{U}_{n})\rightarrow (\mathcal{U}_{m}\cap
\mathcal{U}_{n})\times _{M}G_{m}$ be the restriction of $\chi _{m}^{-1}$ to $%
\pi ^{-1}(\mathcal{U}_{n}\cap \mathcal{U}_{m})$. The composite
diffeomorphism $\Lambda _{nm}\in G$%
\begin{equation}
\Lambda _{mn}:(\mathcal{U}_{n}\cap \mathcal{U}_{m})\times G_{n}\rightarrow (%
\mathcal{U}_{m}\cap \mathcal{U}_{n})\times _{M}G_{m}\text{,}
\end{equation}%
defined as%
\begin{equation}
\Lambda _{ij}\left( x\right) \equiv \rho _{ji}\circ \rho _{ij}^{-1}=\chi _{i%
\text{, }x}\circ \chi _{j\text{, }x}^{-1}:\mathbb{F}\rightarrow \mathbb{F}
\end{equation}%
constitute the transition function between bundle charts $\rho _{nm}$ and $%
\rho _{mn}$ ($\circ $ represents the group composition operation) where the
diffeomorphism $\chi _{i\text{, }x}:\mathbb{F}\rightarrow \mathbb{F}_{x}$ is
written as $\chi _{i\text{, }x}(g):=\chi _{i}\left( x\text{, }g\right) $ and
satisfies $\chi _{j}\left( x\text{, }g\right) =\chi _{i}\left( x\text{, }%
\Lambda _{ij}\left( x\right) g\right) $. The transition functions $\left\{
\Lambda _{ij}\right\} $ can be interpreted as passive gauge transformations.
They satisfy the identity $\Lambda _{ii}\left( x\right) $, inverse $\Lambda
_{ij}\left( x\right) =\Lambda _{ji}^{-1}\left( x\right) $ and cocycle $%
\Lambda _{ij}\left( x\right) \Lambda _{jk}\left( x\right) =\Lambda
_{ik}\left( x\right) $ consistency conditions. For trivial bundles, the
transition function reduces to%
\begin{equation}
\Lambda _{ij}\left( x\right) =g_{i}^{-1}g_{j}\text{,}  \label{transition}
\end{equation}%
where $g_{i}:\mathbb{F}\rightarrow \mathbb{F}$ is defined by $g_{i}:=\chi _{i%
\text{, }x}^{-1}\circ \widetilde{\chi }_{i\text{, }x}$ provided the local
trivializations $\left\{ \chi _{i}\right\} $ and $\left\{ \widetilde{\chi }%
_{i}\right\} $ give rise to the same fiber bundle.

A section is defined as a smooth map%
\begin{equation}
s:M\rightarrow \mathbb{P}\text{,}
\end{equation}%
such that $s(x)\in \pi ^{-1}\left( x\right) =\mathbb{F}_{x}$ $\forall x\in M$
and satisfies%
\begin{equation}
\pi \circ s=\left( id\right) _{M}\text{,}
\end{equation}%
where $\left( id\right) _{M}$ is the identity\ element of $M$. It assigns to
each point $x\in M$ a point in the fiber over $x$. Trivial bundles admit
global sections.

A bundle is a principal fiber bundle $\left\langle \mathbb{P}\text{, }%
\mathbb{P}/G\text{, }G\text{, }\pi \right\rangle $ provided the Lie group $G$
acts freely (i.e. if $pg=p$ then $g=\left( id\right) _{G}$) on $\mathbb{P}$
to the right $R_{g}p=pg$, $p\in \mathbb{P}$, preserves fibers on $\mathbb{P}$
($R_{g}:\mathbb{P}\rightarrow \mathbb{P}$), and is transitive on fibers.
Furthermore, there must exist local trivializations compatible with the $G$
action. Hence, $\pi ^{-1}(\mathcal{U}_{i})$ is homeomorphic to $\mathcal{U}%
_{i}\times _{M}G$ and the fibers of $\mathbb{P}$ are diffeomorphic to $G$.
The trivialization or inverse diffeomorphism map is given by%
\begin{equation}
\chi _{i}^{-1}:\pi ^{-1}(\mathcal{U}_{i})\rightarrow \mathcal{U}_{i}\times
_{M}G  \label{trivial}
\end{equation}%
such that $\chi ^{-1}(p)=\left( \pi (p)\text{, }\varphi (p)\right) \in
\mathcal{U}_{i}\times _{M}G$, $p\in \pi ^{-1}(\mathcal{U}_{i})\subset
\mathbb{P}$, where we see from the above definition that $\varphi $ is a
local mapping of $\pi ^{-1}(\mathcal{U}_{i})$ into $G$ satisfying $\varphi
(L_{g}p)$ $=\varphi (p)g$ for any $p\in \pi ^{-1}(\mathcal{U})$ and any $%
g\in G$. Observe that the elements of $\mathbb{P}$ which are projected onto
the same $x\in \left\{ \mathcal{U}_{i}\right\} $ are transformed into one
another by the elements of $G$. In other words, the fibers of $\mathbb{P}$
are the orbits of $G$ and at the same time, the set of elements which are
projected onto the same $x\in \mathcal{U}\subset M$. This observation
motivates calling the action of the group vertical\ and the base manifold
horizontal. The diffeomorphism map $\chi _{i}$ is called the local gauge
since $\chi _{i}^{-1}$ maps $\pi ^{-1}(\mathcal{U}_{i})$\ onto the direct
(Cartesian) product $\mathcal{U}_{i}\times _{M}G$. The action $L_{g}$ of the
structure group $G$ on $\mathbb{P}$ defines an isomorphism of the Lie
algebra $\mathfrak{g}$ of $G$ onto the Lie algebra of vertical vector fields
on $\mathbb{P}$ tangent to the fiber at each $p\in \mathbb{P}$ called
fundamental vector fields%
\begin{equation}
\lambda _{g}:T_{p}\left( \mathbb{P}\right) \rightarrow T_{gp}(\mathbb{P}%
)=T_{\pi (p)}\left( \mathbb{P}\right) \text{,}
\end{equation}%
where $T_{p}\left( \mathbb{P}\right) $ is the space of tangents at $p$, i.e.
$T_{p}\left( \mathbb{P}\right) \in T\left( \mathbb{P}\right) $. The map $%
\lambda $ is a linear isomorphism for every $p\in \mathbb{P}$ and is
invariant with respect to the action of $G$, that is, $\lambda _{g}:\left(
\lambda _{g\ast }T_{p}\left( \mathbb{P}\right) \right) \rightarrow
T_{gp}\left( \mathbb{P}\right) $, where $\lambda _{g\ast }$ is the
differential push forward map induced by $\lambda _{g}$ defined by $\lambda
_{g\ast }:T_{p}\left( \mathbb{P}\right) \rightarrow T_{gp}\left( \mathbb{P}%
\right) $.

Since the principal bundle $\mathbb{P}\left( M\text{, }G\right) $ is a
differentiable manifold, we can define tangent $T\left( \mathbb{P}\right) $
and cotangent $T^{\ast }\left( \mathbb{P}\right) $ bundles. The tangent
space $T_{p}\left( \mathbb{P}\right) $\ defined at each point $p\in \mathbb{P%
}$ may be decomposed into a vertical $V_{p}\left( \mathbb{P}\right) $ and
horizontal $H_{p}\left( \mathbb{P}\right) $ subspace as $T_{p}\left( \mathbb{%
P}\right) :=V_{p}\left( \mathbb{P}\right) \oplus H_{p}\left( \mathbb{P}%
\right) $ (where $\oplus $ represents the direct sum). The space $%
V_{p}\left( \mathbb{P}\right) $ is a subspace of $T_{p}\left( \mathbb{P}%
\right) $ consisting of all tangent vectors to the fiber passing through $%
p\in \mathbb{P}$, and $H_{p}\left( \mathbb{P}\right) $\ is the subspace
complementary to $V_{p}\left( \mathbb{P}\right) $ at $p$. The vertical
subspace $V_{p}\left( \mathbb{P}\right) :=\left\{ X\in T\left( \mathbb{P}%
\right) |\pi \left( X\right) \in \mathcal{U}_{i}\subset M\right\} $ is
uniquely determined by the structure of $\mathbb{P}$, whereas the horizontal
subspace $H_{p}\left( \mathbb{P}\right) $ cannot be uniquely specified. Thus
we require the following condition: when $p$ transforms as $p\rightarrow
p^{\prime }=pg$, $H_{p}\left( \mathbb{P}\right) $ transforms as \cite%
{Nakahara},%
\begin{equation}
R_{g\ast }H_{p}\left( \mathbb{P}\right) \rightarrow H_{p^{\prime }}\left(
\mathbb{P}\right) =R_{g}H_{p}\left( \mathbb{P}\right) =H_{pg}\left( \mathbb{P%
}\right) .
\end{equation}%
Let the local coordinates of $\mathbb{P}\left( M\text{, }G\right) $ be $%
p=\left( x\text{, }g\right) $ where $x\in M$ and $g\in G$. Let $\mathbf{G}%
_{A}$ denote the generators of the Lie algebra $\mathfrak{g}$ corresponding
to group $G$ satisfying the commutators $\left[ \mathbf{G}_{A}\text{, }%
\mathbf{G}_{B}\right] =f_{AB}^{\text{ \ \ \ }C}\mathbf{G}_{C}$, where $%
f_{AB}^{\text{ \ \ \ }C}$ are the structure constants of $G$. Let $\Omega $
be a connection form defined by $\Omega ^{A}:=\Omega _{i}^{A}dx^{i}\in
\mathfrak{g}$. Let $\omega $ be a connection 1-form defined by%
\begin{equation}
\omega :=\widetilde{g}^{-1}\pi _{\mathbb{P}M}^{\ast }\Omega \widetilde{g}+%
\widetilde{g}^{-1}d\widetilde{g}
\end{equation}%
($\ast $ represents the differential pullback map) belonging to $\mathfrak{g}%
\otimes T_{p}^{\ast }\left( \mathbb{P}\right) $ where $T_{p}^{\ast }\left(
\mathbb{P}\right) $ is the space dual to $T_{p}\left( \mathbb{P}\right) $.
The differential pullback map applied to a test function $\varphi $ and $p$%
-forms $\alpha $ and $\beta $ satisfy $f^{\ast }\varphi =\varphi \circ f$, $%
\left( g\circ f\right) ^{\ast }=f^{\ast }g^{\ast }$ and$\ f^{\ast }\left(
\alpha \wedge \beta \right) =f^{\ast }\alpha \wedge f^{\ast }\beta $. If $G$
is represented by a $d$-dimensional $d\times d$ matrix, then $\mathbf{G}_{A}=%
\left[ \mathbf{G}_{\alpha \beta }\right] $,\ $\widetilde{g}=\left[
\widetilde{g}^{\alpha \beta }\right] $, where $\alpha $, $\beta =1$, $2$, $3$%
,$...d$. Thus, $\omega $ assumes the form%
\begin{equation}
\omega _{\alpha }^{\text{ }\beta }=\left( \widetilde{g}^{-1}\right) _{\alpha
\gamma }d\widetilde{g}^{\gamma \beta }+\left( \widetilde{g}^{-1}\right)
_{\rho \gamma }\pi _{\mathbb{P}M}^{\ast }\Omega _{\text{ }\sigma i}^{\rho }%
\mathbf{G}_{\alpha }^{\text{ }\gamma }\widetilde{g}^{\sigma \beta }\otimes
dx^{i}\text{.}
\end{equation}

If $M$ is $n$-dimensional, the tangent space $T_{p}\left( \mathbb{P}\right) $
is $\left( n+d\right) $-dimensional. Since the vertical subspace $%
V_{p}\left( \mathbb{P}\right) $ is tangential to the fiber $G$, it is $d$%
-dimensional. Accordingly, $H_{p}\left( \mathbb{P}\right) $ is $n$%
-dimensional. The basis of $V_{p}\left( \mathbb{P}\right) $ can be taken to
be $\partial _{\alpha \beta }:=\frac{\partial }{\partial g^{\alpha \beta }}$%
. Now, let the basis of $H_{p}\left( \mathbb{P}\right) $ be denoted by%
\begin{equation}
E_{i}:=\partial _{i}+\Gamma _{i}^{\alpha \beta }\partial _{\alpha \beta }%
\text{,}\ i=1\text{, }2\text{, }3,..n\ \text{and}\ \alpha \text{, }\beta =1%
\text{, }2\text{, }3,..d
\end{equation}%
where $\partial _{i}=\frac{\partial }{\partial x^{i}}$. The connection
1-form $\omega $ projects $T_{p}\left( \mathbb{P}\right) $ onto $V_{p}\left(
\mathbb{P}\right) $. In order for $X\in T_{p}\left( \mathbb{P}\right) $ to
belong to $H_{p}\left( \mathbb{P}\right) $, that is for $X\in H_{p}\left(
\mathbb{P}\right) $, $\omega _{p}\left( X\right) =\left\langle \omega \left(
p\right) |X\right\rangle =0$. In other words,
\begin{equation}
H_{p}\left( \mathbb{P}\right) :=\left\{ X\in T_{p}\left( \mathbb{P}\right)
|\omega _{p}\left( X\right) =0\right\} \text{,}
\end{equation}%
from which $\Omega _{i}^{\alpha \beta }$ can be determined. The inner
product appearing in $\omega _{p}\left( X\right) =\left\langle \omega \left(
p\right) |X\right\rangle =0$ is a map $\left\langle \cdot |\cdot
\right\rangle :T_{p}^{\ast }\left( \mathbb{P}\right) \times T_{p}\left(
\mathbb{P}\right) \rightarrow
%TCIMACRO{\U{211d} }%
%BeginExpansion
\mathbb{R}
%EndExpansion
$ defined by $\left\langle W|V\right\rangle =W_{\mu }V^{\nu }\left\langle
dx^{\mu }|\frac{\partial }{\partial x^{\nu }}\right\rangle =W_{\mu }V^{\nu
}\delta _{\nu }^{\mu }$, where the 1-form $W$ and vector $V$ are given by $%
W=W_{\mu }dx^{\mu }$ and $V=V^{\mu }\frac{\partial }{\partial x^{\nu }}$.
Observe also that, $\left\langle dg^{\alpha \beta }|\partial _{\rho \sigma
}\right\rangle =\delta _{\rho }^{\alpha }\delta _{\sigma }^{\beta }$.

We parameterize an arbitrary group element $\widetilde{g}_{\lambda }$ as $%
\widetilde{g}\left( \lambda \right) =e^{\lambda ^{A}\mathbf{G}%
_{A}}=e^{\lambda \cdot \mathbf{G}}$,\ $A=1$,$..dim\left( \mathfrak{g}\right)
$. The right action $R_{\widetilde{g}\left( \lambda \right) }=R_{\exp \left(
\lambda \cdot G\right) }$ on $p\in \mathbb{P}$, i.e. $R_{\exp \left( \lambda
\cdot \mathbf{G}\right) }p=p\exp \left( \lambda \cdot \mathbf{G}\right) $,
defines a curve through $p$ in $\mathbb{P}$. Define a vector $G^{\#}\in
T_{p}\left( \mathbb{P}\right) $ by \cite{Nakahara}%
\begin{equation}
G^{\#}f\left( p\right) :=\frac{d}{dt}f\left( p\exp \left( \lambda \cdot
\mathbf{G}\right) \right) |_{\lambda =0}
\end{equation}%
where $f:\mathbb{P}\rightarrow
%TCIMACRO{\U{211d} }%
%BeginExpansion
\mathbb{R}
%EndExpansion
$ is an arbitrary smooth function. Since the vector $G^{\#}$ is tangent to $%
\mathbb{P}$ at $p$, $G^{\#}\in V_{p}\left( \mathbb{P}\right) $, the
components of the vector $G^{\#}$ are the fundamental vector fields at $p$
which constitute $V(\mathbb{P})$. The components of $G^{\#}$ may also be
viewed as a basis element of the Lie algebra $\mathfrak{g}$. Given $%
G^{\#}\in V_{p}\left( \mathbb{P}\right) $, $\mathbf{G}\in \mathfrak{g}$,%
\begin{eqnarray}
\omega _{p}\left( G^{\#}\right) &=&\left\langle \omega \left( p\right)
|G^{\#}\right\rangle =\widetilde{g}^{-1}d\widetilde{g}\left( G^{\#}\right) +%
\widetilde{g}^{-1}\pi _{\mathbb{P}M}^{\ast }\Omega \widetilde{g}\left(
G^{\#}\right)  \notag \\
&=&\widetilde{g}_{p}^{-1}\widetilde{g}_{p}\frac{d}{d\lambda }\left( \exp
\left( \lambda \cdot \mathbf{G}\right) \right) |_{\lambda =0}\text{,}
\end{eqnarray}%
where use was made of $\pi _{\mathbb{P}M\ast }G^{\#}=0$. Hence, $\omega
_{p}\left( G^{\#}\right) =\mathbf{G}$. An arbitrary vector $X\in H_{p}\left(
\mathbb{P}\right) $ may be expanded in a basis spanning $H_{p}\left( \mathbb{%
P}\right) $ as $X:=\beta ^{i}E_{i}$. By direct computation, one can show%
\begin{equation}
\left\langle \omega _{\alpha }^{\text{ }\beta }|X\right\rangle =\left(
\widetilde{g}^{-1}\right) _{\alpha \gamma }\beta ^{i}\Gamma _{i}^{\gamma
\beta }+\left( \widetilde{g}^{-1}\right) _{\alpha \gamma }\pi _{\mathbb{P}%
M}^{\ast }\Omega _{\text{ }\sigma i}^{\rho }\beta ^{i}\mathbf{G}_{\rho
}^{\gamma }\widetilde{g}^{\sigma \beta }=0\text{, }\forall \beta ^{i}
\label{conn-vect}
\end{equation}%
Equation (\ref{conn-vect}) yields%
\begin{equation}
\left( \widetilde{g}^{-1}\right) _{\alpha \gamma }\Gamma _{i}^{\gamma \beta
}+\left( \widetilde{g}^{-1}\right) _{\alpha \gamma }\pi _{\mathbb{P}M}^{\ast
}\Omega _{\text{ }\sigma i}^{\rho }\mathbf{G}_{\rho }^{\gamma }\widetilde{g}%
^{\sigma \beta }=0\text{,}
\end{equation}%
from which we obtain%
\begin{equation}
\Gamma _{i}^{\gamma \beta }=-\pi _{\mathbb{P}M}^{\ast }\Omega _{\text{ }%
\sigma i}^{\rho }\mathbf{G}_{\rho }^{\gamma }\widetilde{g}^{\sigma \beta }%
\text{.}
\end{equation}%
In this manner, the horizontal component is completely determined. An
arbitrary tangent vector $\mathfrak{X}\in T_{p}\left( \mathbb{P}\right) $
defined at $p\in \mathbb{P}$ takes the form%
\begin{equation}
\mathfrak{X}=A^{\alpha \beta }\partial _{\alpha \beta }+B^{i}\left( \partial
_{i}-\pi _{\mathbb{P}M}^{\ast }\Omega _{\text{ }\sigma i}^{\rho }\mathbf{G}%
_{\rho }^{\alpha }\widetilde{g}^{\sigma \beta }\partial _{\alpha \beta
}\right) ,
\end{equation}%
where $A^{\alpha \beta }$ and $B^{i}$ are constants. The vector field $%
\mathfrak{X}$ is comprised of horizontal $\mathfrak{X}^{H}:=B^{i}\left(
\partial _{i}-\pi _{\mathbb{P}M}^{\ast }\Omega _{\text{ }\sigma i}^{\rho }%
\mathbf{G}_{\rho }^{\alpha }\widetilde{g}^{\sigma \beta }\partial _{\alpha
\beta }\right) \in H\left( \mathbb{P}\right) $ and vertical $\mathfrak{X}%
^{V}:=A^{\alpha \beta }\partial _{\alpha \beta }\in V\left( \mathbb{P}%
\right) $ components.

Let $\mathfrak{X}\in T_{p}\left( \mathbb{P}\right) $ and $g\in \mathbf{G}$,\
then%
\begin{equation}
R_{g}^{\ast }\omega \left( \mathfrak{X}\right) =\omega \left( R_{g\ast }%
\mathfrak{X}\right) =\widetilde{g}_{pg}^{-1}\Omega \left( R_{g\ast }%
\mathfrak{X}\right) \widetilde{g}_{pg}+\widetilde{g}_{pg}^{-1}d\widetilde{g}%
_{pg}\left( R_{g\ast }\mathfrak{X}\right) \text{,}  \label{Rightw}
\end{equation}%
Observing that $\widetilde{g}_{pg}=\widetilde{g}_{p}g$ and $\widetilde{g}%
_{gp}^{-1}=g^{-1}\widetilde{g}_{p}^{-1}$ the first term on the RHS of (\ref%
{Rightw}) reduces to $\widetilde{g}_{pg}^{-1}\Omega \left( R_{g\ast }%
\mathfrak{X}\right) \widetilde{g}_{pg}=g^{-1}\widetilde{g}_{p}^{-1}\Omega
\left( R_{g\ast }\mathfrak{X}\right) \widetilde{g}_{p}g$ while the second
term gives $\widetilde{g}_{pg}^{-1}d\widetilde{g}_{pg}\left( R_{g\ast }%
\mathfrak{X}\right) =g^{-1}\widetilde{g}_{p}^{-1}d\left( R_{g\ast }\mathfrak{%
X}\right) \widetilde{g}_{p}g$. We therefore conclude%
\begin{equation}
R_{g}^{\ast }\omega _{\lambda }=ad_{g^{-1}}\omega _{\lambda }\text{,}
\end{equation}%
where the adjoint map $ad$ is defined by%
\begin{equation}
ad_{g}Y:=L_{g\ast }\circ R_{g^{-1}\ast }\circ Y=gYg^{-1}\text{, \ }%
ad_{g^{-1}}Y:=g^{-1}Yg\text{.}
\end{equation}

The potential $\Omega ^{A}$ can be obtained from $\omega $ as $\Omega
^{A}=s^{\ast }\omega $. To demonstrate this, let $Y\in T_{p}\left( M\right) $
and $\widetilde{g}$ be specified by the inverse diffeomorphism or
trivialization map (\ref{trivial}) with $\chi _{\lambda }^{-1}\left(
p\right) =\left( x\text{, }\widetilde{g}_{\lambda }\right) $ for $p\left(
x\right) =s_{\lambda }\left( x\right) \cdot \widetilde{g}_{\lambda }$. We
find $s_{i}^{\ast }\omega \left( Y\right) =\widetilde{g}^{-1}\Omega \left(
\pi _{\ast }s_{i\ast }Y\right) \widetilde{g}+\widetilde{g}^{-1}d\widetilde{g}%
\left( s_{i\ast }Y\right) $, where we \cite{Nakahara} have used $s_{i\ast
}Y\in T_{s_{i}}\left( \mathbb{P}\right) $, $\pi _{\ast }s_{i\ast }=\left(
id\right) _{T_{p}\left( M\right) }$ and $\widetilde{g}=\left( id\right) _{G}$
at $s_{i}$ implying $\widetilde{g}^{-1}d\widetilde{g}\left( s_{i\ast
}Y\right) =0$. Hence,%
\begin{equation}
s_{i}^{\ast }\omega \left( Y\right) =\Omega \left( Y\right) \text{.}
\end{equation}

To determine the gauge transformation of the connection 1-form $\omega $ we
use the fact that $R_{\widetilde{g}\ast }X=X\widetilde{g}$ for $X\in
T_{p}\left( M\right) $ and the transition functions $\widetilde{g}_{nm}\in G$
defined between neighboring bundle charts (\ref{transition}). By direct
computation we get%
\begin{eqnarray}
c_{j\ast }X &=&\frac{d}{dt}c_{j}\left( \lambda \left( t\right) \right)
|_{t=0}=\frac{d}{dt}\left[ c_{i}\left( \lambda \left( t\right) \right) \cdot
\widetilde{g}_{ij}\right] |_{t=0}  \notag \\
&=&R_{\widetilde{g}_{ij}\ast }c_{i}^{\ast }\left( X\right) +\left(
\widetilde{g}_{ji}^{-1}\left( x\right) d\widetilde{g}_{ij}\left( X\right)
\right) ^{\#}\text{.}
\end{eqnarray}%
where $\lambda \left( t\right) $ is a curve in $M$ with boundary values $%
\lambda \left( 0\right) =m$ and $\frac{d}{dt}\lambda \left( t\right)
|_{t=0}=X$. Thus, we obtain the useful result%
\begin{equation}
c_{\ast }X=R_{\widetilde{g}\ast }\left( c_{\ast }X\right) +\left( \widetilde{%
g}^{-1}d\widetilde{g}\left( X\right) \right) ^{\#}\text{.}  \label{trans1}
\end{equation}%
Applying $\omega $ to (\ref{trans1}) we get%
\begin{equation}
\omega \left( c_{\ast }X\right) =c^{\ast }\omega \left( X\right) =ad_{%
\widetilde{g}^{-1}}c^{\ast }\omega \left( X\right) +\widetilde{g}^{-1}d%
\widetilde{g}\left( X\right) \text{, }\forall X\text{.}
\end{equation}%
Hence, the gauge transformation of the local gauge potential $\Omega $ reads,%
\begin{equation}
\Omega \rightarrow \Omega ^{\prime }=ad_{\widetilde{g}^{-1}}\left( d+\Omega
\right) =\widetilde{g}^{-1}\left( d+\Omega \right) \widetilde{g}\text{.}
\label{Lconn-trans}
\end{equation}%
Since $\Omega =c^{\ast }\omega $ we obtain from (\ref{Lconn-trans}) the
gauge transformation law of $\omega $%
\begin{equation}
\omega \rightarrow \omega ^{\prime }=\widetilde{g}^{-1}\left( d+\omega
\right) \widetilde{g}\text{.}
\end{equation}
Now we are ready to undertake the task to construct the bundle
structure of gravitation.

\section{The Generalized Bundle Structure of Gravitation}

Let us recall the definition of gauge transformations in the
context of ordinary fiber bundles. This step will be extremely
relevant to induce metric and dynamics from invariance properties.
Given a principal fiber bundle $\mathbb{P}(M$, $G$; $%
\pi )$ with base space $M$ and standard $G$-diffeomorphic fiber, gauge
transformations are characterized by bundle isomorphisms \cite{Giachetta} $%
\lambda :\mathbb{P}\rightarrow \mathbb{P}$ exhausting all diffeomorphisms $%
\lambda _{M}$ on $M$. This mapping is called an automorphism of $\mathbb{P}$
provided it is equivariant with respect to the action of $G$. This amounts
to restricting the action $\lambda $ of $G$ along local fibers leaving the
base space unaffected. Indeed, with regard to gauge theories of internal
symmetry groups, a gauge transformation is a fiber preserving bundle
automorphism, i.e. diffeomorphisms $\lambda $ with $\lambda _{M}=\left(
id\right) _{M}$. The automorphisms $\lambda $ form a group called the
automorphism group $Aut_{\mathbb{P}}$ of $\mathbb{P}$. The gauge
transformations form a subgroup of $Aut_{\mathbb{P}}$ called the gauge group
$G\left( Aut_{\mathbb{P}}\right) $ (or $G$ in short) of $\mathbb{P}$.

The map $\lambda $ is required to satisfy two conditions, namely its
commutability with the right action of $G$ $[$the equivariance condition $%
\lambda \left( R_{g}(p)\right) =\lambda \left( pg\right) =\lambda \left(
p\right) g]$%
\begin{equation}
\lambda \circ R_{g}(p)=R_{g}(p)\circ \lambda \text{, \ }p\in \mathbb{P}\text{%
, }g\in G
\end{equation}%
according to which fibers are mapped into fibers, and the verticality
condition%
\begin{equation}
\pi \circ \lambda \left( u\right) =\pi \left( u\right) \text{,}
\end{equation}%
where $u$ and $\lambda \left( u\right) $ belong to the same fiber. The last
condition ensures that no diffeomorphisms $\lambda _{M}:M\rightarrow M$
given by%
\begin{equation}
\lambda _{M}\circ \pi \left( u\right) =\pi \circ \lambda \left( u\right)
\text{,}
\end{equation}%
be allowed on the base space $M$. In a gauge description of gravitation, one
is interested in gauging external transformation groups. That is to say the
group action on spacetime coordinates cannot be neglected. The spaces of
internal fiber and external base must be interlocked in the sense that
transformations in one space must induce corresponding transformations in
the other. The usual definition of a gauge transformation, i.e. as a
displacement along local fibers not affecting the base space, must be
generalized to reflect this interlocking. One possible way of framing this
interlocking is to employ a nonlinear realization of the gauge group $G$,
provided a closed subgroup $H\subset G$ exist. The interlocking requirement
is then transformed into the interplay between groups $G$ and one of its
closed subgroups $H$.

Denote by $G$ a Lie group with elements $\left\{ g\right\} $. Let $H$ be a
closed subgroup of $G$ specified by $[37$, $67]$%
\begin{equation}
H:=\left\{ h\in G|\Pi \left( R_{h}g\right) =\pi \left( g\right) \text{, }%
\forall g\in G\right\} \text{,}
\end{equation}%
with elements $\left\{ h\right\} $ and known linear representations $\rho
\left( h\right) $. Here $\Pi $ is the first of the two projection maps in (%
\ref{comp-pro}), and $R_{h}$ is the right group action. Let $M$ be a
differentiable manifold with points $\left\{ x\right\} $ to which $G$ and $H$
may be referred, i.e. $g=g(x)$ and $h=h(x)$. Being that $G$ and $H$ are Lie
groups, they are also manifolds. The right action of $H$ on $G$ induce a
complete partition of $G$ into mutually disjoint orbits $gH$. Since $g=g(x)$%
, all elements of $gH=\left\{ gh_{1}\text{, }gh_{2}\text{, }gh_{3}\text{,}%
\cdot \cdot \cdot \text{, }gh_{n}\right\} $ are defined over the same $x$.
Thus, each orbit $gH$ constitute an equivalence class of point $x$, with
equivalence relation $g\equiv g^{\prime }$ where\ $g^{\prime }=R_{h}g=gh$.
By projecting each equivalence class onto a single element of the quotient
space $\mathcal{M}:=G/H$, the group $G$ becomes organized as a fiber bundle
\ in\ the sense that $G=\bigcup\nolimits_{i}\left\{ g_{i}H\right\} $. In
this manner the manifold $G$ is viewed as a fiber bundle $G\left( \mathcal{M}%
\text{, }H\text{; }\Pi \right) $ with $H$-diffeomorphic fibers $\Pi
^{-1}\left( \xi \right) :G\rightarrow \mathcal{M}=gH$ and base space $%
\mathcal{M}$. A composite principal fiber bundle\textit{\ }$\mathbb{P}(M$, $G
$; $\pi )$ is one whose $G$-diffeomorphic fibers possess the fibered
structure $G\left( \mathcal{M}\text{, }H\text{; }\Pi \right) \simeq \mathcal{%
M}\times $ $H$ described above. The bundle $\mathbb{P}$ is then locally
isomorphic to $M\times G\left( \mathcal{M}\text{, }H\right) $. Moreover,
since an element $g\in G$ is locally homeomorphic to $\mathcal{M}\times H$
the elements of $\mathbb{P}$ are - by transitivity - also locally
homeomorphic to $M\times \mathcal{M}\times H\simeq \Sigma \times H$ where
(locally) $\Sigma \simeq M\times \mathcal{M}$. Thus, an alternative view
\cite{Tresguerres} of $\mathbb{P}(M$, $G$; $\pi )$ is provided by the $%
\mathbb{P}$-associated $H$-bundle $\mathbb{P}(\Sigma $, $H$; $\widetilde{\pi
})$. The total space $\mathbb{P}$ may be regarded as $G\left( \mathcal{M}%
\text{, }H\text{; }\Pi \right) $-bundles over base space $M$ or equivalently
as $H$-fibers attached to manifold $\Sigma \simeq M\times \mathcal{M}$.

The nonlinear realization  technique \cite{CCWZ1, CCWZ2} provides
a way to determine the transformation properties of fields defined
on the quotient
space $G/H$. The nonlinear realization of Diff$\left( 4\text{, }%
%TCIMACRO{\U{211d} }%
%BeginExpansion
\mathbb{R}
%EndExpansion
\right) $ becomes tractable due to a theorem given by V. I. Ogievetsky.
According to the Ogievetsky theorem \cite{BorisovOgievetskii}, the algebra
of the infinite dimensional group Diff$\left( 4\text{, }%
%TCIMACRO{\U{211d} }%
%BeginExpansion
\mathbb{R}
%EndExpansion
\right) $ can be taken as the closure of the finite dimensional algebras of $%
SO(4$, $2)$ and $A(4$, $%
%TCIMACRO{\U{211d} }%
%BeginExpansion
\mathbb{R}
%EndExpansion
)$. Remind that the Lorentz group generates transformations that
preserve the quadratic form on Minkowski spacetime built from the
metric tensor, while the special conformal group generates
infinitesimal angle-preserving transformations on Minkowski
spacetime. The affine group is a generalization of the
Poincar\'{e} group where the Lorentz group is replaced by the
group of general linear transformations \cite{felix}. As such, the
affine group generates translations, Lorentz transformations,
volume preserving shear and volume
changing dilation transformations. As a consequence, the nonlinear realization of Diff$\left( 4%
\text{, }%
%TCIMACRO{\U{211d} }%
%BeginExpansion
\mathbb{R}
%EndExpansion
\right) /SO(3$, $1)$ can be constructed by taking a simultaneous realization
of the conformal group $SO(4$, $2)$ and the affine group $A(4$, $%
%TCIMACRO{\U{211d} }%
%BeginExpansion
\mathbb{R}
%EndExpansion
):=%
%TCIMACRO{\U{211d} }%
%BeginExpansion
\mathbb{R}
%EndExpansion
^{4}\rtimes GL(4$, $%
%TCIMACRO{\U{211d} }%
%BeginExpansion
\mathbb{R}
%EndExpansion
)$ on the coset spaces $A(4$, $%
%TCIMACRO{\U{211d} }%
%BeginExpansion
\mathbb{R}
%EndExpansion
)/SO(3$, $1)$ and $SO(4$, $2)/SO(3$, $1)$. One possible interpretation of
this theorem is that the conform-affine group (defined below) may be the
largest subgroup of Diff$\left( 4\text{, }%
%TCIMACRO{\U{211d} }%
%BeginExpansion
\mathbb{R}
%EndExpansion
\right) $ whose transformations may be put into the form of a
generalized coordinate transformation. We remark that a nonlinear
realization  can be made linear by embedding the representation in
a sufficiently higher dimensional space. Alternatively, a linear
group realization becomes nonlinear when subject to constraints.
One type of relevant constraints may be those responsible for
symmetry reduction from Diff$\left( 4\text{, }%
%TCIMACRO{\U{211d} }%
%BeginExpansion
\mathbb{R}
%EndExpansion
\right) $ to $SO(3$, $1)$ for instance.

We take the group $CA(3$, $1)$ as the basic symmetry group $G$.
The conformal-affine group
consists of the groups $SO(4$, $2)$ and $A(4$, $%
%TCIMACRO{\U{211d} }%
%BeginExpansion
\mathbb{R}
%EndExpansion
)$. In particular, conformal-affine is proportional to the union $SO(4$, $2)\cup A(4$, $%
%TCIMACRO{\U{211d} }%
%BeginExpansion
\mathbb{R}
%EndExpansion
)$. We know however (see section \textit{Conform-Affine Lie Algebra}) that
the affine and special conformal groups have several group generators in
common. These common generators reside in the intersection $SO(4$, $2)\cap
A(4$, $%
%TCIMACRO{\U{211d} }%
%BeginExpansion
\mathbb{R}
%EndExpansion
)$ of the two groups, within which there are \textit{two copies}
of $\Pi :=D\times P(3$, $1)$, where $D$ is the group of scale
transformations (dilations)\ and $P(3$, $1):=T\left( 3\text{,
}1\right) \rtimes SO(3$, $1)$ is the Poincar\'{e} group. We define
the conformal-affine group as the union of the affine
and conformal groups minus \textit{one copy} of the overlap $\Pi $, i.e. $%
CA(3$, $1):=SO(4$, $2)\cup A(4$, $%
%TCIMACRO{\U{211d} }%
%BeginExpansion
\mathbb{R}
%EndExpansion
)-\Pi $. Being defined in this way we recognize that $CA(3$, $1)$ is a $24$
parameter Lie group representing the action of Lorentz transformations $(6)$%
, translations $(4)$, special conformal transformations $(4)$,
spacetime shears $(9)$ and scale transformations $(1)$. In this
paper, we are obtaining the nonlinear realization of $CA(3$, $1)$
modulo $SO(3$, $1)$.

\section{The Conformal-Affine Lie Algebra}

In order to implement the nonlinear realization procedure, we choose the partition Diff$(4$, $%
%TCIMACRO{\U{211d} }%
%BeginExpansion
\mathbb{R}
%EndExpansion
)$ with respect to the Lorentz group. By Ogievetsky's theorem \cite%
{BorisovOgievetskii}, we identify representations of Diff$(4$, $%
%TCIMACRO{\U{211d} }%
%BeginExpansion
\mathbb{R}
%EndExpansion
)/SO(3$, $1)$ with those of $CA(3$, $1)/SO(3$, $1)$. The $20$ generators of
affine transformations can be decomposed into the $4$ translational $\mathbf{%
P}_{\mu }^{\text{Aff}}$ and $16$ $GL(4$, $%
%TCIMACRO{\U{211d} }%
%BeginExpansion
\mathbb{R}
%EndExpansion
)$ transformations $\mathbf{\Lambda }_{\alpha }^{\text{ }\beta }$. The $16$
generators $\mathbf{\Lambda }_{\alpha }^{\text{ }\beta }$ may be further
decomposed into the $6$ Lorentz generators $\mathbf{L}_{\alpha }^{\text{ }%
\beta }$ plus the remaining $10$ generators of symmetric linear
transformation $\mathbf{S}_{\alpha }^{\text{ }\beta }$, that is, $\mathbf{%
\Lambda }_{\text{ }\beta }^{\alpha }=\mathbf{L}_{\text{ }\beta }^{\alpha }+%
\mathbf{S}_{\text{ }\beta }^{\alpha }$. The $10$ parameter symmetric linear
generators $\mathbf{S}_{\alpha }^{\text{ }\beta }$ can be factored into the $%
9$ parameter shear (the traceless part of $\mathbf{S}_{\alpha }^{\text{ }%
\beta }$) generator defined by $^{\dagger }\mathbf{S}_{\alpha }^{\text{ }%
\beta }=\mathbf{S}_{\alpha }^{\text{ }\beta }-\frac{1}{4}\delta _{\alpha }^{%
\text{ }\beta }\mathbf{D}$, and the $1$ parameter dilaton generator $\mathbf{%
D}=tr\left( \mathbf{S}_{\alpha }^{\text{ }\beta }\right) $. Shear
transformations generated by $^{\dagger }\mathbf{S}_{\alpha }^{\text{ }\beta
}$ describe shape changing, volume preserving deformations, while the
dilaton generator gives rise to volume changing transformations. The four
diagonal elements of $\mathbf{S}_{\alpha }^{\text{ }\beta }$ correspond to
the generators of projective transformations. The $15$ generators of
conformal transformations are defined in terms of the set $\left\{
J_{AB}\right\} $ where $A=0$, $1$, $2$,..$5$. The elements $J_{AB}$ can be
decomposed into translations $\mathbf{P}_{\mu }^{\text{Conf}}:=J_{5\mu
}+J_{6\mu }$, special conformal generators $\mathbf{\Delta }_{\mu }:=J_{5\mu
}-J_{6\mu }$, dilatons $\mathbf{D}:=J_{56}$ and the Lorentz generators $%
\mathbf{L}_{\alpha \beta }:=J_{\alpha \beta }$. The Lie algebra of $CA(3$, $%
1)$ is characterized by the commutation relations%
\begin{equation}
\begin{array}{c}
\left[ \mathbf{\Lambda }_{\alpha \beta }\text{, }\mathbf{D}\right] =\left[
\mathbf{\Delta }_{\alpha }\text{, }\mathbf{\Delta }_{\beta }\right] =0\text{%
, }\left[ \mathbf{P}_{\alpha }\text{, }\mathbf{P}_{\beta }\right] =\left[
\mathbf{D}\text{, }\mathbf{D}\right] =0\text{,} \\
\left[ \mathbf{L}_{\alpha \beta }\text{, }\mathbf{P}_{\mu }\right] =io_{\mu
\lbrack \alpha }\mathbf{P}_{\beta ]}\text{, }\left[ \mathbf{L}_{\alpha \beta
}\text{, }\mathbf{\Delta }_{\gamma }\right] =io_{[\alpha |\gamma }\mathbf{%
\Delta }_{|\beta ]}\text{,} \\
\left[ \mathbf{\Lambda }_{\text{ }\beta }^{\alpha }\text{, }\mathbf{P}_{\mu }%
\right] =i\delta _{\mu }^{\alpha }\mathbf{P}_{\beta }\text{, }\left[ \mathbf{%
\Lambda }_{\text{ }\beta }^{\alpha }\text{, }\mathbf{\Delta }_{\mu }\right]
=i\delta _{\mu }^{\alpha }\mathbf{\Delta }_{\beta }\text{,} \\
\left[ \mathbf{S}_{\alpha \beta }\text{, }\mathbf{P}_{\mu }\right] =io_{\mu
(\alpha }\mathbf{P}_{\beta )}\text{, }\left[ \mathbf{P}_{\alpha }\text{, }%
\mathbf{D}\right] =-i\mathbf{P}_{\alpha }\text{,} \\
\left[ \mathbf{L}_{\alpha \beta }\text{, }\mathbf{L}_{\mu \nu }\right]
=-i\left( o_{\alpha \lbrack \mu }\mathbf{L}_{\nu ]\beta }-o_{\beta \lbrack
\mu }\mathbf{L}_{\nu ]\alpha }\right) \text{,} \\
\left[ \mathbf{S}_{\alpha \beta }\text{, }\mathbf{S}_{\mu \nu }\right]
=i\left( o_{\alpha (\mu }\mathbf{L}_{\nu )\beta }-o_{\beta (\mu }\mathbf{L}%
_{\nu )\alpha }\right) \text{,} \\
\left[ \mathbf{L}_{\alpha \beta }\text{, }\mathbf{S}_{\mu \nu }\right]
=i\left( o_{\alpha (\mu }\mathbf{S}_{\nu )\beta }-o_{\beta (\mu }\mathbf{S}%
_{\nu )\alpha }\right) \text{,} \\
\left[ \mathbf{\Delta }_{\alpha }\text{, }\mathbf{D}\right] =i\mathbf{\Delta
}_{\alpha }\text{, }\left[ \mathbf{S}_{\mu \nu }\text{, }\mathbf{\Delta }%
_{\alpha }\right] =io_{\alpha (\mu }\mathbf{\Delta }_{\nu )}\text{,} \\
\left[ \mathbf{\Lambda }_{\text{ }\beta }^{\alpha }\text{, }\mathbf{\Lambda }%
_{\text{ }\nu }^{\mu }\right] =i\left( \delta _{\nu }^{\alpha }\mathbf{%
\Lambda }_{\text{ }\beta }^{\mu }-\delta _{\beta }^{\mu }\mathbf{\Lambda }_{%
\text{ }\nu }^{\alpha }\right) \text{,} \\
\left[ \mathbf{P}_{\alpha }\text{, }\mathbf{\Delta }_{\beta }\right]
=2i\left( o_{\alpha \beta }\mathbf{D}-\mathbf{L}_{\alpha \beta }\right)
\text{, }%
\end{array}%
\end{equation}%
where $o_{\alpha \beta }=diag\left( -1\text{, }1\text{, }1\text{,
}1\right) $ is Lorentz group metric. The above algebra is the core
of the nonlinear realization and, in some sense, of the Invariance
Induced Gravity.

\section{Group Actions and Bundle Morphisms}

Let us now introduce the main ingredients required to specify the
structure of the fiber bundle, namely the canonical projection,
the sections, etc.  We follow the prescription in
\cite{Tresguerres} for constructing the composite fiber bundle,
but implement the program for the conformal-affine group.

The composite bundle \textit{\ }$\mathbb{P}(\Sigma $, $H$;
$\widetilde{\pi })$
is comprised of $H$-fibers, base space $\Sigma \left( M\text{, }\mathcal{M}%
\right) $ and a composite map
\begin{equation}
\widetilde{\pi }\overset{\text{def}}{=}\widetilde{\pi }_{\Sigma M}\circ \Pi
_{\mathbb{P}\Sigma }:\mathbb{P}\rightarrow \Sigma \rightarrow M\text{,}
\end{equation}%
with component projections%
\begin{equation}
\Pi _{\mathbb{P}\Sigma }:\mathbb{P}\rightarrow \Sigma \text{, \ }\widetilde{%
\pi }_{\Sigma M}:\Sigma \rightarrow M\text{.}  \label{comp-pro}
\end{equation}%
The projection $\Pi _{\mathbb{P}\Sigma }$ maps the point $\left( p\in
\mathbb{P}\text{, }R_{h}p\in \mathbb{P}\right) $ into point $\left( x\text{,
}\xi \right) \in $ $\Sigma $. There is a correspondence between sections $%
s_{M\Sigma }:M\rightarrow \Sigma $ and the projection $\Pi _{\mathbb{P}%
\Sigma }:\mathbb{P}\rightarrow \Sigma $ in the sense that both maps project
their functional argument onto elements of $\Sigma $. This is formalized by
the relation, $\Pi _{\mathbb{P}\Sigma }\left( p\right) =s_{M\Sigma }\circ
\pi _{\mathbb{P}M}\left( p\right) $. Hence, the total projection is given by%
\begin{equation}
\widetilde{\pi }:=\pi _{\mathbb{P}M}=\widetilde{\pi }_{\Sigma M}\circ \Pi _{%
\mathbb{P}\Sigma }.
\end{equation}%
Associated with the projections $\widetilde{\pi }_{\Sigma M}$ and $\Pi _{%
\mathbb{P}\Sigma }$ are the corresponding local sections%
\begin{equation}
s_{M\Sigma }:\mathcal{U}\rightarrow \widetilde{\pi }_{\Sigma M}^{-1}\left(
\mathcal{U}\right) \subset \Sigma \text{, }s_{\Sigma \mathbb{P}}:\mathcal{V}%
\rightarrow \Pi _{\mathbb{P}\Sigma }^{-1}\left( \mathcal{V}\right) \subset
\mathbb{P}\text{,}
\end{equation}%
with neighborhoods $\mathcal{U}\subset M$ and $\mathcal{V}\subset \Sigma $
satisfying%
\begin{equation}
\widetilde{\pi }_{\Sigma M}\circ s_{M\Sigma }=\left( id\right) _{M}\text{, }%
\Pi _{\mathbb{P}\Sigma }\circ s_{\Sigma \mathbb{P}}=\left( id\right)
_{\Sigma }\text{.}
\end{equation}%
The bundle injection $\widetilde{\pi }^{-1}\left( \mathcal{U}\right) $ is
the inverse image of $\widetilde{\pi }\left( \mathcal{U}\right) $ and is
called the fiber over $\mathcal{U}$. The equivalence class $R_{h}p=pH\in
\widetilde{\pi }_{\Sigma M}^{-1}\left( \mathcal{U}\right) $ of left cosets
is the fiber of $\mathbb{P}\left( \Sigma \text{, }H\right) $ while each
orbit $pH$ through $p\in \mathbb{P}$\ projects into a single element $Q\in $
$\Sigma $. In analogy to the total bundle projection (\ref{comp-pro}), a
total section of $\mathbb{P}$ is given by the total section composition%
\begin{equation}
s_{M\mathbb{P}}=s_{\Sigma \mathbb{P}}\circ s_{M\Sigma }.  \label{comp-sect}
\end{equation}%
Let elements of $G/H$ be labeled by the parameter $\xi $. Functions on $G/H$
are represented by continuous coset functions $c(\xi )$ parameterized by $%
\xi $. These elements are referred to as cosets to the right of $H$ with
respect to $g\in G$. Indeed, the orbits of the right action of $H$ on $G$
are the left cosets $R_{h}g=gH$. For a given section $s_{M\mathbb{P}}\left(
x\in M\right) \in \pi _{\mathbb{P}M}^{-1}$ with local coordinates $\left( x%
\text{, }g\right) $ one can perform decompositions of the partial fibers $%
s_{M\Sigma }$ and $s_{\Sigma \mathbb{P}}$ as:
\begin{equation}
s_{M\Sigma }\left( x\right) =\widetilde{c}_{M\Sigma }\left( x\right) \cdot
c=R_{c^{\prime }}\circ \widetilde{c}_{M\Sigma }\left( x\right) \text{; }%
c=c\left( \xi \right) \text{,}  \label{sect-Msig}
\end{equation}%
\begin{equation}
s_{\Sigma \mathbb{P}}\left( x\text{, }\xi \right) =\widetilde{c}_{\Sigma
\mathbb{P}}\left( x\text{, }\xi \right) \cdot a^{\prime }=R_{a^{\prime
}}\circ \widetilde{c}_{\Sigma \mathbb{P}}\left( x\text{, }\xi \right) \text{%
; }a^{\prime }\in H\text{,}  \label{sect-sigP}
\end{equation}%
with the null sections $\left\{ \widetilde{c}_{M\Sigma }\left( x\right)
\right\} $ and $\left\{ \widetilde{c}_{\Sigma \mathbb{P}}\left( x\text{, }%
\xi \right) \right\} $ having coordinates $\left( x\text{, }\left( id\right)
_{\mathcal{M}}\right) $ and $\left( x\text{, }\xi \text{, }\left( id\right)
_{H}\right) $ respectively. A null or zero section is a map that sends every
point $x\in M$ to the origin of the fiber $\pi ^{-1}\left( x\right) $ over $%
x $, i.e. $\chi _{i}^{-1}\left( \widetilde{c}\left( x\right) \right) =\left(
x\text{, }0\right) $ in any trivialization. The trivialization map $\chi
_{i}^{-1}$ is defined in (\ref{trivial}) The identity map appearing in the
above trivializations are defined as $\left( id\right) _{\mathcal{M}}:%
\mathcal{M}\rightarrow \mathcal{M}$ and\ $\left( id\right) _{H}:H\rightarrow
H$. We assume the total null bundle section be given by the composition law
\begin{equation}
\widetilde{c}_{M\mathbb{P}}=\widetilde{c}_{\Sigma \mathbb{P}}\circ
\widetilde{c}_{M\Sigma }\text{.}  \label{comp-nullsect}
\end{equation}%
The images of two sections $s_{\Sigma \mathbb{P}}$ and $s_{M\Sigma }$ over $%
x\in M$ must coincide, implying $s_{\Sigma \mathbb{P}}\left(
x\text{, }\xi \right) =s_{M\Sigma }\left( x\right) $. Using
(\ref{comp-sect}) with (\ref{sect-Msig}), (\ref{sect-sigP}) and
(\ref{comp-nullsect}), we arrive at the total bundle section
decomposition
\begin{equation}
s_{M\mathbb{P}}\left( x\right) =\widetilde{c}_{M\mathbb{P}}\left( x\right)
\cdot g=R_{g}\circ \widetilde{c}_{M\mathbb{P}}\left( x\right)
\label{sMP-null}
\end{equation}%
provided $g=c\cdot a$ and
\begin{equation}
\widetilde{c}_{\Sigma \mathbb{P}}=R_{c^{-1}}\circ \widetilde{c}_{\Sigma
\mathbb{P}}\left( x\text{, }\xi \right) \circ R_{c}\text{.}  \label{csigP}
\end{equation}%
The pullback of $\widetilde{c}_{\Sigma \mathbb{P}}$, defined \cite%
{Tresguerres} as
\begin{equation}
\widetilde{c}_{\xi }\left( x\right) =\left( s_{M\Sigma }^{\ast }\widetilde{c}%
_{\Sigma \mathbb{P}}\right) \left( x\right) =\widetilde{c}_{\Sigma \mathbb{P}%
}\circ s_{M\Sigma }=\widetilde{c}_{\Sigma \mathbb{P}}\left( x\text{, }\xi
\right) \text{,}  \label{null-xi}
\end{equation}%
ensures the coincidence of images of sections $\widetilde{c}_{\xi }\left(
x\right) :M\rightarrow \mathbb{P}$ and $\widetilde{c}_{\Sigma \mathbb{P}%
}\left( x\text{, }\xi \right) :\Sigma \rightarrow \mathbb{P}$,
respectively. With the aid of the above results, we arrive at the
equation
\begin{equation}
\widetilde{c}_{\Sigma \mathbb{P}}\left( x\text{, }\xi \right) =\widetilde{c}%
_{M\mathbb{P}}\left( x\right) \cdot c\left( \xi \right) \text{,}
\label{csigP1}
\end{equation}
which will be extremely useful in the following.

\section{Nonlinear Realizations and  Generalized Gauge
Transformations}

The generalized gauge transformation law is obtained by comparing bundle
elements $p\in \mathbb{P}$ that differ by the left action of elements of the
principal group $G$, $L_{g\in G}$. An arbitrary element $p\in \mathbb{P}$
can be written in terms of the null section with the aid of (\ref{sMP-null}%
), (\ref{csigP}) and (\ref{csigP1}) as
\begin{equation}
p=s_{M\mathbb{P}}\left( x\right) =R_{a}\circ \widetilde{c}_{\Sigma \mathbb{P}%
}\left( x\text{, }\xi \right) \text{, }a\in H\text{.}  \label{p-initial}
\end{equation}%
Performing a gauge transformation on $p$, we obtain the orbit
$\lambda \left(
p\right) $ defining a curve through $\left( x\text{, }\xi \right) $ in $%
\Sigma $%
\begin{equation}
\lambda \left( p\right) =L_{g\left( x\right) }\circ p=R_{a^{\prime }}\circ
\widetilde{c}_{\Sigma \mathbb{P}}\left( x\text{, }\xi ^{\prime }\right)
\text{; \ }g\left( x\right) \in G\text{, \ }a^{\prime }\in H\text{.}
\label{p-trans}
\end{equation}%
Comparison of (\ref{p-initial}) with (\ref{p-trans}) leads to%
\begin{equation}
L_{g\left( x\right) }\circ R_{a}\circ \widetilde{c}_{\Sigma \mathbb{P}%
}\left( x\text{, }\xi \right) =R_{a^{\prime }}\circ \widetilde{c}_{\Sigma
\mathbb{P}}\left( x\text{, }\xi ^{\prime }\right) \text{.}  \label{inter}
\end{equation}%
By virtue of the commutability \cite{Nakahara} of left and right group
translations of elements belonging to $G$, i.e. $L_{g}\circ R_{h}=R_{h}\circ
L_{g}$, Eq.(\ref{inter}) may be recast as%
\begin{equation}
L_{g\left( x\right) }\circ \widetilde{c}_{\Sigma \mathbb{P}}\left( x\text{, }%
\xi \right) =R_{h}\circ \widetilde{c}_{\Sigma \mathbb{P}}\left( x\text{, }%
\xi ^{\prime }\right) \text{.}  \label{gen-gauge}
\end{equation}%
where $R_{a^{-1}}\circ R_{a^{\prime }}\equiv R_{a^{\prime }a^{-1}}:=R_{h}$
and\ $a^{\prime }a^{-1}\equiv h\in H$. Equation (\ref{gen-gauge}) constitute
a generalized gauge transformation. Performing the pullback of (\ref%
{gen-gauge}) with respect to the section $s_{M\Sigma }$ leads to%
\begin{equation}
L_{g\left( x\right) }\circ \widetilde{c}_{\xi }\left( x\right) =R_{h\left(
\xi \text{, }g(x)\right) }\circ \widetilde{c}_{\xi ^{\prime }}\left(
x\right) \text{.}  \label{gen-gauge1}
\end{equation}%
Thus, the left action $L_{g}$ of $G$ is a map that acts on $\mathbb{P}$ and $%
\Sigma $. In particular, $L_{g}$ acting on fibers defined as orbits of the
right action describes diffeomorphisms that transforming fibers over $%
\widetilde{c}_{\xi }\left( x\right) $ into the fibers $\widetilde{c}_{\xi
^{\prime }}\left( x\right) $ of $\Sigma $ while simultaneously being
displaced along $H$ fibers via the action of $R_{h}$. Equation (\ref%
{gen-gauge1}) states that nonlinear realizations of $G$ mod $H$ is
determined by the action of an arbitrary element $g\in G$ on the quotient
space $G/H$ transforming one coset into another as
\begin{equation}
L_{g}:G/H\rightarrow G/H\text{, \ }c(\xi )\rightarrow c(\xi ^{\prime })
\end{equation}%
inducing a diffeomorphism $\xi \rightarrow \xi ^{\prime }$ on $G/H$. To
simplify the action induced by (\ref{gen-gauge1}) for calculation purposes
we proceed as follows. Departing from (\ref{null-xi}) and substituting $%
s_{M\Sigma }=R_{c}\circ \widetilde{c}_{M\mathbb{P}}$ we get%
\begin{equation}
\widetilde{c}_{\xi }\left( x\right) =\widetilde{c}_{\Sigma \mathbb{P}}\circ
R_{c}\circ \widetilde{c}_{M\Sigma }\text{.}  \label{null-xi1}
\end{equation}%
Using $\widetilde{c}_{M\mathbb{P}}\circ R_{c}=R_{c}\circ \widetilde{c}_{M%
\mathbb{P}}$, (\ref{null-xi1}) becomes $\widetilde{c}_{\xi }\left( x\right)
=R_{c}\circ \widetilde{c}_{\Sigma \mathbb{P}}\circ \widetilde{c}_{M\Sigma
}=R_{c}\circ \widetilde{c}_{M\mathbb{P}}$, where the last equality follows
from use of $\widetilde{c}_{M\mathbb{P}}=\widetilde{c}_{\Sigma \mathbb{P}%
}\circ \widetilde{c}_{M\Sigma }$. By way of analogy, we assume $\widetilde{c}%
_{\xi ^{\prime }}\left( x\right) \equiv R_{c^{\prime }}\circ \widetilde{c}_{M%
\mathbb{P}}$. Upon substitution of $\widetilde{c}_{\xi ^{\prime }}$ into (%
\ref{gen-gauge1}) we obtain%
\begin{equation}
L_{g}\circ R_{c}\circ \widetilde{c}_{M\mathbb{P}}=R_{h\left( \xi \text{, }%
g\left( x\right) \right) }\circ R_{c^{\prime }}\circ \widetilde{c}_{M\mathbb{%
P}}\text{,}
\end{equation}%
which after implementing the group actions is equivalent to,
\begin{equation}
g\cdot \widetilde{c}_{M\mathbb{P}}\cdot c=\widetilde{c}_{M\mathbb{P}}\cdot
c^{\prime }\cdot h\text{.}  \label{inter1}
\end{equation}%
Operating on (\ref{inter1}) from the left by $\widetilde{c}_{M\mathbb{P}%
}^{-1}$ and making use of $g=\widetilde{c}_{M\mathbb{P}}^{-1}g\widetilde{c}%
_{M\mathbb{P}}$, we get $\left( \widetilde{c}_{M\mathbb{P}}^{-1}\cdot g\cdot
\widetilde{c}_{M\mathbb{P}}\right) \cdot c=c^{\prime }\cdot h$ which leads
to $g\cdot c_{\xi }=c_{\xi ^{\prime }}\cdot h$, or
\begin{equation}
c^{\prime }=g\cdot c\cdot h^{-1}  \label{NLR}
\end{equation}%
in short, where $c\equiv c_{\xi }$ and $c^{\prime }\equiv c_{\xi ^{\prime }}$%
. Observe that the element $h$ is a function whose argument is the couple $%
\left( \xi \text{, }g\left( x\right) \right) $. The transformation rule (\ref%
{NLR}) is in fact the key equation to determine the nonlinear realizations
of $G$ and specifies a unique $H$-valued field $h(\xi $, $g\left( x\right) )$%
\ on $G/H$.

Consider a family of sections $\left\{ \widehat{c}\left( x\text{, }\xi
\right) \right\} $ defined \cite{TiembloTresguerres1} on $\Sigma $ by
\begin{equation}
\widehat{c}\left( x\text{, }\xi \right) :=c\circ \widetilde{c}\left( x\text{%
, }\xi \right) =c\left( \widetilde{c}\left( x\text{, }\xi \right) \right)
\text{.}  \label{null-family}
\end{equation}%
Taking $\Pi _{\mathbb{P}\Sigma }\circ R_{h}\circ $ $\widetilde{c}_{\Sigma
\mathbb{P}}=\Pi _{\mathbb{P}\Sigma }\circ \widetilde{c}_{\Sigma \mathbb{P}%
}=\left( id\right) _{\Sigma }$ into account, we can explicitly exhibit the
fact that the left action $L_{g}$ of $G$ on the null sections $\widetilde{c}%
_{\Sigma \mathbb{P}}:\mathbb{P}\rightarrow \Sigma $ induces an equivalence
relation between differing elements $\widetilde{c}_{\xi }$, $\widetilde{c}%
_{\xi ^{\prime }}\in \Sigma $ given by%
\begin{equation}
\Pi _{\mathbb{P}\Sigma }\circ L_{g}\circ \widehat{c}_{\xi }=\Pi _{\mathbb{P}%
\Sigma }\circ R_{h\left( \xi \text{, }g\left( x\right) \right) }\circ
\widehat{c}_{\xi ^{\prime }}=R_{h\left( \xi \text{, }g\left( x\right)
\right) }\circ \widetilde{c}_{\xi ^{\prime }}\text{,}
\end{equation}%
so that%
\begin{equation}
\widetilde{c}_{\xi }^{\prime }:=R_{h\left( \xi \text{, }g\left( x\right)
\right) }\circ \widetilde{c}_{\xi ^{\prime }}=L_{g}\circ \widetilde{c}_{\xi }%
\text{.}  \label{inter2}
\end{equation}%
From (\ref{inter2}) we can write%
\begin{equation}
\widetilde{c}_{\xi }\overset{L_{g}}{\longmapsto }\widetilde{c}_{\xi
}^{\prime }=R_{h\left( \xi \text{, }g\left( x\right) \right) }\circ
\widetilde{c}_{\xi ^{\prime }}\text{ }\forall h\in H\text{.}  \label{inter3}
\end{equation}%
Equation (\ref{inter3}) gives rise to a complete partition of $G/H$ into
equivalence classes $\Pi _{\mathbb{P}\Sigma }^{-1}\left( \xi \right) $ of
left cosets \cite{TiembloTresguerres1, TiembloTresguerres3}%
\begin{equation}
cH=\left\{ R_{h\left( \xi \text{, }g\left( x\right) \right) }\circ c/c\in G/H%
\text{, }\forall h\in H\right\} =\left\{ ch_{1}\text{, }ch_{2}\text{,..., }%
ch_{n}\right\} \text{,}
\end{equation}%
where $c\in (G-H)$\ plays the role of the fibers attached to each point of $%
\Sigma $. The elements $ch_{i}$\ are single representatives of
each equivalence class $R_{h\left( \xi \text{, }g\left( x\right)
\right) }\circ c=cH\in \widetilde{\pi }_{\Sigma M}^{-1}\left(
\mathcal{U}\right) $. Thus, any diffeomorphism $L_{g}\circ
\widetilde{c}_{\xi }$ on $\Sigma $ together with the $H$-valued
function $h\left( \xi \text{, }g\left( x\right) \right) $
determine a unique gauge transformation $\widetilde{c}_{\xi
}^{\prime }=R_{h\left( \xi \text{, }g\left( x\right) \right)
}\circ \widetilde{c}_{\xi ^{\prime }}$.\ This demonstrates that
gauge transformations are those diffeomorphisms on $\Sigma $ that
map fibers over $c\left( \xi \right) $ into fibers over $c\left(
\xi ^{\prime }\right) $ and simultaneously preserve the action of
$H$.

\section{The Covariant Coset Field Transformations}

We now proceed to determine the transformation behavior of
parameters belonging to $G/H$. The elements of the
conformal-affine and Lorentz groups are
respectively parameterized about the identity element as%
\begin{equation}
g=e^{i\epsilon ^{\alpha }\mathbf{P}_{\alpha }}e^{i\alpha ^{\mu \nu }\text{ }%
^{\dagger }\mathbf{S}_{\mu \nu }}e^{i\beta ^{\mu \nu }\mathbf{L}_{\mu \nu
}}e^{ib^{\alpha }\mathbf{\Delta }_{\alpha }}e^{i\varphi \mathbf{D}}\text{,}\
h=e^{iu^{\mu \nu }\mathbf{L}_{\mu \nu }}\text{.}
\end{equation}%
Elements of the coset space $G/H$ are coordinatized by
\begin{equation}
c=e^{-i\xi ^{\alpha }\mathbf{P}_{\alpha }}e^{ih^{\mu \nu }\text{ }^{\dagger }%
\mathbf{S}_{\mu \nu }}e^{i\zeta ^{\alpha }\mathbf{\Delta }_{\alpha
}}e^{i\phi \mathbf{D}}\text{.}
\end{equation}%
We consider transformations with infinitesimal group parameters $\epsilon
^{\alpha }$, $\alpha ^{\mu \nu }$, $\beta ^{\mu \nu }$, $b^{\alpha }$ and $%
\varphi $. The transformed coset parameters read $\xi ^{\prime \alpha }=\xi
^{\alpha }+\delta \xi ^{\alpha }$, $h^{\prime \mu \nu }=h^{\mu \nu }+\delta
h^{\mu \nu }$, $\zeta ^{\prime \alpha }=\zeta ^{\alpha }+\delta \zeta
^{\alpha }$ and $\phi ^{\prime }=\phi +\delta \phi $. Note that $u^{\mu \nu
} $ is infinitesimal. The translational coset field variations reads%
\begin{equation}
\delta \xi ^{\alpha }=-\left( \alpha _{\beta }^{\text{ \ }\alpha }+\beta
_{\beta }^{\text{ \ }\alpha }\right) \xi ^{\beta }-\epsilon ^{\alpha
}-\varphi \xi ^{\alpha }-\left[ \left\vert \xi \right\vert ^{2}b^{\alpha
}-2\left( b\cdot \xi \right) \xi ^{\alpha }\right] \text{.}
\end{equation}%
For the dilatons we get,%
\begin{equation}
\delta \phi =\varphi +2\left( b\cdot \xi \right) -\left\{ u_{\text{ }\beta
}^{\alpha }\xi ^{\beta }+\epsilon ^{\alpha }+\varphi \xi ^{\alpha }+\left[
b^{\alpha }\left\vert \xi \right\vert ^{2}-2\left( b\cdot \xi \right) \xi
^{\alpha }\right] \right\} \partial _{\alpha }\phi \text{.}
\end{equation}%
Similarly for the special conformal $4$-boosts we find,%
\begin{eqnarray}
\delta \zeta ^{\alpha } &=&u_{\text{ }\beta }^{\alpha }\zeta ^{\beta
}+b^{\alpha }-\varphi \zeta ^{\alpha }+2\left[ \left( b\cdot \xi \right)
\zeta ^{\alpha }-\left( b\cdot \zeta \right) \xi ^{\alpha }\right] + \\
&&  \notag \\
&&-\left\{ u_{\text{ }\lambda }^{\beta }\xi ^{\lambda }+\epsilon ^{\beta
}+\varphi \xi ^{\beta }+\left[ b^{\beta }\left\vert \xi \right\vert
^{2}-2\left( b\cdot \xi \right) \xi ^{\beta }\right] \right\} \partial
_{\beta }\zeta ^{\alpha }\text{.}  \notag
\end{eqnarray}%
Observe the homogeneous part of the special conformal coset parameter $\zeta
^{\alpha }$ has the same structure as that of the translational parameter $%
\xi ^{\alpha }$ (with the substitutions: $\zeta ^{\alpha }\rightarrow -\xi
^{\alpha }$ and $-\epsilon ^{\alpha }\rightarrow b^{\alpha }$). For the
shear parameters we obtain%
\begin{equation}
\delta r^{\alpha \beta }=\left( \alpha ^{\gamma \alpha }+\beta ^{\gamma
\alpha }\right) r_{\gamma }^{\text{ \ }\beta }+u_{\text{ }\gamma }^{\beta
}r^{\alpha \gamma }+2b^{[\alpha }\xi ^{\rho ]}r_{\rho }^{\text{ \ }\beta }%
\text{,}
\end{equation}%
where $r^{\alpha \beta }:=e^{h^{\alpha \beta }}$. From $\delta r^{\alpha
\beta }$\ we obtain the nonlinear Lorentz transformation%
\begin{equation}
u^{\alpha \beta }=\beta ^{\alpha \beta }+2b^{[\alpha }\xi ^{\beta ]}-\alpha
^{\mu \nu }\tanh \left\{ \frac{1}{2}\ln \left[ r_{\text{ }\mu }^{\alpha
}\left( r^{-1}\right) _{\text{ }\nu }^{\beta }\right] \right\} \text{.}
\label{NL-lorentz}
\end{equation}%
In the limit of vanishing special conformal $4$-boost, this result coincides
with that of Pinto \textit{et al.} \cite{Lopez-Pinto}. For vanishing shear,
the result of Julve \textit{et al} \cite{Julve} is obtained.

In this section, all covariant coset field transformations have
been  determined directly from the nonlinear transformation law
(\ref{NLR}). We observe that the translational coset parameter
transforms as a coordinate under the action of $G$. From the shear
coset variation, the explicit form of the nonlinear Lorentz-like
transformation was obtained. From (\ref{NL-lorentz}) it is clear
that $u^{\alpha \beta }$ contains the linear Lorentz parameter
in addition to conformal and shear contributions via the nonlinear $4$%
-boosts and symmetric $GL_{4}$ parameters.

\section{The Decomposition of  Connections in $\protect\pi _{\mathbb{P}M}:\mathbb{%
P}\rightarrow M$ into components in $\protect\pi _{\mathbb{P}\Sigma }:%
\mathbb{P}\rightarrow \Sigma $ and in $\protect\pi _{\Sigma
M}:\Sigma \rightarrow M$}

At this point, it is useful to classify all possible
decompositions of connections in order to achieve all the
conformal-affine nonlinear gauge potentials. This step is
essential to have all the ingredients to construct the induced
metric and dynamics related to the conformal-affine group.

Depending on which bundle is considered, either the total bundle $\mathbb{P}%
\rightarrow M$ or the intermediate bundles $\mathbb{P}\rightarrow \Sigma $, $%
\Sigma \rightarrow M$, we may construct corresponding Ehresmann connections
for the respective space. With respect to $M$, we have the connection form%
\begin{equation}
\omega =\widetilde{g}^{-1}\left( d+\pi _{\mathbb{P}M}^{\ast }\Omega
_{M}\right) \widetilde{g}\text{.}  \label{wPM}
\end{equation}%
The gauge potential $\Omega _{M}$ is defined in the standard manner as the
pullback of the connection $\omega $ by the null section $\widetilde{c}_{M%
\mathbb{P}}$, $\Omega _{M}=\widetilde{c}_{M\mathbb{P}}^{\ast }\omega \in
T^{\ast }\left( M\right) $. With regard to the space $\Sigma $ an
alternative form of the connection is given by%
\begin{equation}
\omega =a^{-1}\left( d+\pi _{\mathbb{P}\Sigma }^{\ast }\Gamma _{\Sigma
}\right) a\text{,}  \label{wPsig}
\end{equation}%
where the connection on $\Sigma $ reads $\Gamma _{\Sigma }=\widetilde{c}%
_{\Sigma \mathbb{P}}^{\ast }\omega $. Carrying out a similar
analysis  and evaluating the tangent vector $X\in T_{p}\left(
\Sigma \right) $ at each point $\xi $ along the curve $c_{\xi }$
on the coset space
$G/H$ that coincides with the section $\widetilde{c}_{\Sigma \mathbb{P}%
}^{\ast }$, we find the gauge transformation law%
\begin{equation}
\omega \rightarrow \omega ^{\prime }=ad_{h^{-1}}\left( d+\omega \right)
\text{.}
\end{equation}%
Comparison of (\ref{wPM}) and \ref{wPsig} leads to $\pi _{\mathbb{P}\Sigma
}^{\ast }\Gamma _{\Sigma }=c^{-1}\left( d+\pi _{\mathbb{P}M}^{\ast }\Omega
_{M}\right) c$. Taking account of $\widetilde{c}_{\Sigma \mathbb{P}}^{\ast
}\Pi _{\mathbb{P}\Sigma }^{\ast }=\left( id\right) _{T^{\ast }\left( \Sigma
\right) }$ which follows from $\Pi _{\mathbb{P}\Sigma }\circ \widetilde{c}%
_{\Sigma \mathbb{P}}=\left( id\right) _{\Sigma }$, we deduce
\begin{equation}
\Gamma _{\Sigma }=\widetilde{c}_{\Sigma \mathbb{P}}^{\ast }\left[
c^{-1}\left( d+\pi _{\mathbb{P}M}^{\ast }\Omega _{M}\right) c\right] \text{.}
\end{equation}%
By use of the family of sections pulled back to $\Sigma $ introduced in (\ref%
{null-family}) we find $\widetilde{c}_{\Sigma \mathbb{P}}^{\ast }\left(
c^{-1}dc\right) =\widehat{c}$ $^{-1}d\widehat{c}$ and\ $\widetilde{c}%
_{\Sigma \mathbb{P}}^{\ast }R_{c}^{\ast }=R_{\widehat{c}}^{\ast }\widetilde{c%
}_{\Sigma \mathbb{P}}^{\ast }$. Recalling $\widetilde{\pi }_{\mathbb{P}%
M}^{\ast }=\widetilde{\pi }_{\mathbb{P}\Sigma }^{\ast }\widetilde{\pi }%
_{\Sigma M}^{\ast }$, we get $c^{-1}\widetilde{\pi }_{\mathbb{P}M}^{\ast
}\Omega _{M}c=R_{c}^{\ast }\widetilde{\pi }_{\mathbb{P}M}^{\ast }\Omega _{M}$%
. With these results in hand, we obtain the alternative form of the
connection $\Gamma _{\Sigma }$,%
\begin{equation}
\Gamma _{\Sigma }=\widehat{c}^{-1}\left( d+\pi _{\Sigma M}^{\ast }\Omega
_{M}\right) \widehat{c}\text{.}
\end{equation}%
Completing the pullback of $\Gamma _{\Sigma }$ to $M$ by means of $%
\widetilde{c}_{M\Sigma }$ we obtain, $\Gamma _{M}=\widetilde{c}_{M\Sigma
}^{\ast }\Gamma _{\Sigma }$. By use of $\Gamma _{\Sigma }=\widetilde{c}%
_{\Sigma \mathbb{P}}^{\ast }\omega $ and (\ref{null-xi}) we find $\Gamma
_{M}=s_{M\Sigma }^{\ast }\widetilde{c}_{\Sigma \mathbb{P}}^{\ast }\omega =%
\widetilde{c}_{\xi }^{\ast }\omega $. In terms of the substitution $\widehat{%
c}\left( x\text{, }\xi \right) \rightarrow \overline{c}\left( x\right) $
where $\overline{c}\left( x\right) $ is the pullback of $\widehat{c}\left( x%
\text{, }\xi \right) $ to $M$ defined as $\overline{c}\left( x\right)
=s_{M\Sigma }^{\ast }\widehat{c}=c\left( \widetilde{c}_{\xi }\left( x\right)
\right) $, we arrive at the desired result%
\begin{equation}
\mathbf{\Gamma }\equiv \Gamma _{M}=\overline{c}^{-1}\left( d+\Omega
_{M}\right) \overline{c}\text{,}
\end{equation}%
which explicitly relates the connection $\mathbf{\Gamma }$ on $\Sigma $
pulled back to $M$ to its counterpart $\Omega _{M}$.

The gauge transformation behavior of $\mathbf{\Gamma }$\ may be determined
directly by use of (\ref{Lconn-trans}) and the transformation $\widetilde{c}%
^{\prime }=g\widetilde{c}h^{-1}$. We calculate%
\begin{equation}
\mathbf{\Gamma }^{\prime }=h\widetilde{c}^{-1}g^{-1}d\left( g\widetilde{c}%
h^{-1}\right) +h\widetilde{c}^{-1}\Omega \widetilde{c}h^{-1}+h\widetilde{c}%
^{-1}\left( dg^{-1}\right) g\widetilde{c}h^{-1}\text{.}
\end{equation}%
Observing however, that
\begin{equation}
h\widetilde{c}^{-1}g^{-1}d\left( g\widetilde{c}h^{-1}\right) =h\widetilde{c}%
^{-1}\left( g^{-1}dg\right) \widetilde{c}h^{-1}+h\widetilde{c}^{-1}d%
\widetilde{c}h^{-1}+hdh^{-1}\text{,}
\end{equation}%
we obtain%
\begin{equation}
\mathbf{\Gamma }^{\prime }=h\left[ \widetilde{c}^{-1}\left( d+\Omega \right)
\widetilde{c}\right] h^{-1}+hdh^{-1}+h\widetilde{c}^{-1}d\left(
gg^{-1}\right) \widetilde{c}h^{-1}\text{.}
\end{equation}%
Thus, we arrive at the gauge transformation law
\begin{equation}
\mathbf{\Gamma }^{\prime }=h\mathbf{\Gamma }h^{-1}+hdh^{-1}\text{.}
\label{NLR-transf}
\end{equation}%
According to the Lie algebra decomposition of $\mathfrak{g}$ into $\mathfrak{%
h}$ and $\mathfrak{c}$, the connection $\Gamma _{\Sigma }$ can be divided
into $\mathbf{\Gamma }_{H}$ defined on the subgroup $H$ and $\mathbf{\Gamma }%
_{G/H}$ defined on $G/H$. From the transformation law (\ref{NLR-transf}) it
is clear that $\mathbf{\Gamma }_{H}$ transforms inhomogeneously
\begin{equation}
\mathbf{\Gamma }_{H}^{\prime }=h\mathbf{\Gamma }_{H}h^{-1}+hdh^{-1}\text{,}
\end{equation}%
while $\Gamma _{G/H}$ transforms as a tensor
\begin{equation}
\mathbf{\Gamma }_{G/H}^{\prime }=h\mathbf{\Gamma }_{G/H}h^{-1}\text{.}
\end{equation}%
In this regard, only $\Gamma _{H}$ transforms as a true connection. We use
the gauge potential $\mathbf{\Gamma }$ to define the gauge covariant
derivative%
\begin{equation}
\mathbf{\nabla }:=\left( d+\rho \left( \mathbf{\Gamma }\right) \right)
\end{equation}%
acting on $\psi $ as $\nabla \psi =\left( d+\rho \left( \Gamma \right)
\right) \psi $ with the desired transformation property%
\begin{equation}
\left( \nabla \psi \left( c(\xi )\right) \right) ^{\prime }=\rho \left(
h(\xi \text{, }g)\right) \nabla \psi \left( c(\xi )\right) \simeq \left(
1+iu\left( \xi \text{, }g\right) \rho \left( H\right) \right) \nabla \psi
\left( c(\xi )\right)
\end{equation}%
leading to%
\begin{equation}
\delta \left( \nabla \psi \left( c(\xi )\right) \right) =iu\left( \xi \text{%
, }g\right) \rho \left( H\right) \nabla \psi \left( c(\xi )\right) \text{.}
\end{equation}
Let us now classify the conformal-affine gauge potentials
considering the various components of the decomposition.

\subsection{Conformal-Affine Nonlinear Gauge Potential in $\protect\pi _{%
\mathbb{P}M}:\mathbb{P\rightarrow }$ $M$}

The ordinary gauge potential defined on the total base space $M$ reads%
\begin{equation}
\Omega =-i\left( \overset{\text{T}}{\Gamma }\text{ }^{\alpha }\mathbf{P}%
_{\alpha }+\overset{\text{C}}{\Gamma }\text{ }^{\alpha }\mathbf{\Delta }%
_{\alpha }+\overset{\text{D}}{\Gamma }\mathbf{D}+\overset{\text{GL}}{\Gamma }%
\text{ }^{\alpha \beta }\text{ }^{\dagger }\mathbf{\Lambda }_{\alpha \beta
}\right) \text{.}  \label{omega}
\end{equation}%
The horizontal basis vectors that span the horizontal tangent space $\mathbb{%
H}(\mathbb{P})$ of $\pi _{\mathbb{P}M}:\mathbb{P\rightarrow }M$ are given by%
\begin{equation}
E_{i}=\widetilde{c}_{M\mathbb{P\ast }}\partial _{i}-\Omega _{i}\text{.}
\end{equation}%
The explicit form of the connections (\ref{omega}) are given by%
\begin{equation}
\omega =-i\left[ V_{M}^{\mu }\widetilde{\chi }_{\mu }^{\text{ }\nu }\mathbf{P%
}_{\nu }-i\left( i\overline{\Theta }_{\left( ^{\dagger }\mathbf{\Lambda }%
\right) }^{\alpha \beta }+\widetilde{\pi }_{\mathbb{P}M}^{\ast }\overset{%
\text{GL}}{\Gamma }\text{ }^{\alpha \beta }\right) \widetilde{\chi }_{\alpha
}^{\text{ }\nu }\widetilde{\chi }_{\beta }^{\text{ }\nu }\text{ }^{\dagger }%
\mathbf{\Lambda }_{\mu \nu }+\vartheta _{M}^{\mu }\widetilde{\beta }_{\mu }^{%
\text{ }\nu }\mathbf{\Delta }_{\nu }-i\widetilde{\pi }_{\mathbb{P}M}^{\ast
}\Phi _{M}\mathbf{D}\right]
\end{equation}%
where $\overline{\Theta }_{\left( ^{\dagger }\Lambda \right) }^{\alpha \beta
}=\overline{\Theta }_{\left( \mathbf{L}\right) }^{\alpha \beta }+\overline{%
\Theta }_{\left( \text{SY}\right) }^{\alpha \beta }$, with right invariant
Maurer-Cartan forms%
\begin{equation}
\overline{\Theta }_{\left( \mathbf{L}\right) }^{\mu \nu }=i\widetilde{\beta }%
_{\text{ \ \ \ \ }\gamma }^{[\nu |}d\widetilde{\beta }^{|\mu ]\gamma
}-2idb^{\mu }\epsilon ^{\nu }\text{ and }\overline{\Theta }_{\left( \text{SY}%
\right) }^{\mu \nu }=i\widetilde{\alpha }_{\text{ \ \ \ \ }\gamma }^{(\nu |}d%
\widetilde{\alpha }^{|\mu )\gamma }\text{.}
\end{equation}%
The linear connection $\Omega _{M}$ varies under the action of $G$ as%
\begin{equation}
\delta \Omega =\Omega ^{\prime }-\Omega =\delta \overset{\text{T}}{\Gamma }%
\text{ }^{\mu }\mathbf{P}_{\mu }+\delta \overset{\text{C}}{\Gamma }\text{ }%
^{\mu }\mathbf{\Delta }_{\mu }+\delta \overset{\text{D}}{\Gamma }\mathbf{D}%
+\delta \overset{\text{GL}}{\Gamma }\text{ }^{\beta \nu }\text{ }^{\dagger }%
\mathbf{\Lambda }_{\beta \nu }
\end{equation}%
where%
\begin{equation}
\begin{array}{c}
\delta \overset{\text{T}}{\Gamma }\text{ }^{\mu }=\text{ }^{\dagger }\overset%
{\text{GL}}{D}\epsilon ^{\mu }-\overset{\text{T}}{\Gamma }\text{ }^{\alpha
}\left( \alpha _{\alpha }^{\text{ }\mu }+\beta _{\alpha }^{\text{ }\mu
}+\varphi \delta _{\alpha }^{\text{ }\mu }\right) -\overset{\text{D}}{\Gamma
}\epsilon ^{\mu }\text{,} \\
\\
\delta \overset{\text{C}}{\Gamma }\text{ }^{\mu }=\text{ }^{\dagger }\overset%
{\text{GL}}{D}b^{\mu }-\overset{\text{C}}{\Gamma }\text{ }^{\alpha }\left(
\alpha _{\alpha }^{\text{ }\mu }+\beta _{\alpha }^{\text{ }\mu }-\varphi
\delta _{\alpha }^{\text{ }\mu }\right) +\overset{\text{D}}{\Gamma }b^{\mu }%
\text{,} \\
\\
\delta \overset{\text{GL}}{\Gamma }\text{ }^{\alpha \beta }=\text{ }%
^{\dagger }\overset{\text{GL}}{D}\left( \alpha ^{\alpha \beta }+\beta
^{\alpha \beta }\right) +\left( \overset{\text{T}}{\Gamma }\text{ }^{[\alpha
}b^{\beta ]}+\overset{\text{C}}{\Gamma }\text{ }^{[\alpha }\epsilon ^{\beta
]}\right) \text{,} \\
\\
\delta \overset{\text{D}}{\Gamma }=d\varphi +2\left( \overset{\text{C}}{%
\Gamma }\text{ }^{\alpha }\epsilon _{\alpha }-\overset{\text{T}}{\Gamma }%
\text{ }^{\alpha }b_{\alpha }\right) \text{.}%
\end{array}%
\end{equation}%
The components of $\overline{\omega }$ on $M$ are identified as spacetime
quantities and are determined from the pullback of the corresponding
(quotient space) quantities defined on $\Sigma $:%
\begin{equation}
V_{M}^{\mu }=s_{M\Sigma }^{\ast }V_{\Sigma }^{\mu }\text{,}\ \vartheta
_{M}^{\mu }=s_{M\Sigma }^{\ast }\vartheta _{\Sigma }^{\mu }\text{, }\Phi
_{M}=s_{M\Sigma }^{\ast }\Phi _{\Sigma }\ \text{and}\ \Gamma _{M}^{\mu \nu
}=s_{M\Sigma }^{\ast }\Gamma _{\Sigma }^{\mu \nu }\text{.}
\end{equation}%
In the following, we depart from the alternative form of the connection $%
\omega =a^{-1}\left( d+\Pi _{\mathbb{P}\Sigma }^{\ast }\Gamma _{\Sigma
}\right) a$, $\forall $ $a\in H$ on $\Sigma $.

\subsection{Conformal-Affine Nonlinear Gauge Potential in $\protect\pi _{%
\mathbb{P}\Sigma }:\mathbb{P}\rightarrow \Sigma $}

The components of $\omega $ in $\mathbb{P}\rightarrow \Sigma $ are oriented
along the Lie algebra basis of $H$%
\begin{equation}
\overset{\mathbf{L}}{\omega }=a^{-1}\left( d+i\widetilde{\pi }_{\mathbb{P}%
\Sigma }^{\ast }\overset{\circ }{\Gamma }\text{ }^{\alpha \beta }\mathbf{L}%
_{\alpha \beta }\right) a=-i\overset{\mathbf{L}}{\omega }\text{ }^{\alpha
\beta }\mathbf{L}_{\alpha \beta }\text{,}
\end{equation}%
where%
\begin{equation}
\overset{\mathbf{L}}{\omega }\text{ }^{\alpha \beta }:=\left( i\overline{%
\Theta }_{\left( \mathbf{L}\right) }^{\rho \sigma }+\widetilde{\pi }_{%
\mathbb{P}\Sigma }^{\ast }\Gamma _{\left[ \mathbf{L}\right] }^{\rho \sigma
}\right) \widetilde{\beta }_{[\rho }^{\text{ }\alpha }\widetilde{\beta }%
_{\sigma ]}^{\text{ }\beta }\text{.}
\end{equation}

\subsection{Conformal-Affine Nonlinear Gauge Potential on $\Pi _{\Sigma
M}:\Sigma \rightarrow M$}

The components of $\omega $\ in $\Pi _{\Sigma M}:\Sigma \rightarrow M$ are
oriented \cite{Tresguerres} along the Lie algebra basis of the quotient
space $G/H$ belonging to $\Sigma $%
\begin{eqnarray}
\overset{\mathbf{P}}{\omega } &=&-ia^{-1}\left( \widetilde{\pi }_{\Sigma
M}^{\ast }V_{\Sigma }^{\nu }\mathbf{P}_{\nu }\right) a=-i\overset{\mathbf{P}}%
{\omega }\text{ }^{\mu }\mathbf{P}_{\mu }\text{,} \\
&&  \notag \\
\overset{\mathbf{\Delta }}{\omega } &=&-ia^{-1}\left( \widetilde{\pi }%
_{\Sigma M}^{\ast }\vartheta _{\Sigma }^{\nu }\mathbf{\Delta }_{\nu }\right)
a=-i\overset{\mathbf{\Delta }}{\omega }\text{ }^{\mu }\mathbf{\Delta }_{\mu }%
\text{,} \\
&&  \notag \\
\overset{\mathbf{D}}{\omega } &=&-ia^{-1}\left( \widetilde{\pi }_{\Sigma
M}^{\ast }\Phi _{\Sigma }\mathbf{D}\right) a=-i\omega _{\left[ \mathbf{D}%
\right] }\mathbf{D}\text{,} \\
&&  \notag \\
\overset{\text{SY}}{\omega } &=&-ia^{-1}\left( \widetilde{\pi }_{\Sigma
M}^{\ast }\Upsilon ^{\alpha \beta }\mathbf{S}_{\alpha \beta }\right) a=-i%
\overset{\text{SY}}{\omega }\text{ }^{\alpha \beta }\mathbf{S}_{\alpha \beta
}\text{,}
\end{eqnarray}%
where%
\begin{eqnarray}
\overset{\mathbf{P}}{\omega }\text{ }^{\mu } &:&=\widetilde{\pi }_{\Sigma
M}^{\ast }V_{\Sigma }^{\nu }\widetilde{\beta }_{\nu }^{\text{ }\mu }\text{,}%
\ \overset{\mathbf{\Delta }}{\omega }\text{ }^{\mu }:=\widetilde{\pi }%
_{\Sigma M}^{\ast }\vartheta _{\Sigma }^{\nu }\widetilde{\beta }_{\nu }^{%
\text{ }\mu }\text{,} \\
&&  \notag \\
\omega _{\left[ \mathbf{D}\right] } &:&=\widetilde{\pi }_{\Sigma M}^{\ast
}\Phi _{\Sigma }\text{,}\ \overset{\text{SY}}{\omega }\text{ }^{\alpha \beta
}:=\widetilde{\pi }_{\mathbb{P}\Sigma }^{\ast }\Upsilon ^{\rho \sigma }%
\widetilde{\alpha }_{(\rho }^{\text{ }\alpha }\widetilde{\alpha }_{\sigma
)}^{\text{ }\beta }\text{.}
\end{eqnarray}%
By direct computation we obtain%
\begin{equation}
\mathbf{\Gamma }_{\Sigma }^{\text{CA}}=-i\left( V_{\Sigma }^{\mu }\mathbf{P}%
_{\mu }+i\vartheta _{\Sigma }^{\mu }\mathbf{\Delta }_{\mu }+\Phi _{\Sigma }%
\mathbf{D}+\Gamma _{\Sigma }^{\alpha \beta }\mathbf{\Lambda }_{\alpha \beta
}\right) .
\end{equation}%
The nonlinear translational and special conformal connection coefficients $%
V_{\Sigma }^{\nu }$ and $\vartheta _{\Sigma }^{\nu }$\ read%
\begin{equation}
V_{\Sigma }^{\beta }=\widetilde{\pi }_{\Sigma M}^{\ast }\left[ e^{\phi
}\left( \upsilon ^{\beta }\left( \xi \right) +r_{\text{ }\sigma }^{\alpha }%
\overset{\text{C}}{\Gamma }\text{ }^{\sigma }\mathfrak{B}_{\alpha }^{\text{ }%
\beta }\left( \xi \right) \right) \right] \text{,}  \label{NL-p1}
\end{equation}%
\begin{equation}
\vartheta _{\Sigma }^{\beta }=\widetilde{\pi }_{\Sigma M}^{\ast }\left[
e^{-\phi }\left( \upsilon ^{\beta }\left( \zeta \right) +\upsilon ^{\sigma
}\left( \xi \right) \mathfrak{B}_{\sigma }^{\text{ }\beta }\left( \zeta
\right) \right) \right] \text{,}  \label{NL-p2}
\end{equation}%
with%
\begin{equation}
\upsilon _{i}^{\beta }\left( \xi \right) :=r_{\sigma }^{\beta }\left(
\overset{\text{GL}}{^{\dagger }D_{i}}\xi ^{\sigma }+\overset{\text{D}}{%
\Gamma }_{i}\xi ^{\sigma }+\overset{\text{T}}{\Gamma }\text{ }_{i}^{\sigma
}\right) \text{, }\mathfrak{B}_{\alpha }^{\text{ }\rho }\left( \xi \right)
:=\left( \left\vert \xi \right\vert ^{2}\delta _{\alpha }^{\text{ }\rho
}-2\xi _{\alpha }\xi ^{\rho }\right) \text{.}
\end{equation}%
The nonlinear $GL_{4}$ and dilaton connections are given by%
\begin{equation}
\Gamma _{\Sigma }^{\mu \nu }=\widehat{\Gamma }\text{ }^{\mu \nu }+2\zeta
^{\lbrack \mu }\varpi ^{\nu ]}\text{,}  \label{NL-p3}
\end{equation}%
\begin{equation}
\Phi =\widetilde{\pi }_{\Sigma M}^{\ast }\left( \zeta _{\beta }\varpi
^{\beta }\right) -\frac{1}{2}d\phi \text{,}  \label{NL-p4}
\end{equation}%
with%
\begin{equation}
\widehat{\Gamma }\text{ }^{\mu \nu }:=\widetilde{\pi }_{\Sigma M}^{\ast }%
\left[ \left( r^{-1}\right) _{\;\sigma }^{\mu }\overset{\text{GL}}{\Gamma }%
\text{ }^{\sigma \beta }r_{\beta }^{\;\nu }-\left( r^{-1}\right) _{\;\sigma
}^{\mu }dr^{\sigma \nu }\right]
\end{equation}%
and%
\begin{equation}
\varpi ^{\nu }:=\upsilon ^{\nu }+r_{\text{ }\alpha }^{\nu }\overset{\text{C}}%
{\Gamma }\text{ }^{\alpha }\text{.}
\end{equation}%
The nonlinear $GL_{4}$ connection can be expanded in the $GL_{4}$ Lie
algebra according to $\Gamma ^{\alpha \beta }$ $^{\dagger }\mathbf{\Lambda }%
_{\alpha \beta }=\overset{\circ }{\Gamma }$ $^{\alpha \beta }\mathbf{L}%
_{\alpha \beta }+\Upsilon ^{\alpha \beta }$ $^{\dagger }\mathbf{S}_{\alpha
\beta }$, where%
\begin{equation}
\overset{\circ }{\Gamma }\text{ }_{\Sigma }^{\alpha \beta }:=\widehat{\Gamma
}\text{ }^{[\alpha \beta ]}+2\zeta ^{\lbrack \alpha }\varpi ^{\beta ]}\text{%
, }\Upsilon _{\Sigma }^{\alpha \beta }:=\widehat{\Gamma }\text{ }^{(\alpha
\beta )}\text{.}  \label{rest}
\end{equation}%
The symmetric $GL_{4}$ (shear) gauge fields $\Upsilon $ are distortion
fields describing the difference between the general linear connection and
the Levi-Civita connection.

We define the (group) algebra bases $e_{\nu }$ and $h_{\nu }$ dual to the
translational and special conformal 1-forms $V^{\mu }$ and $\vartheta ^{\mu
} $ as%
\begin{eqnarray}
e_{\mu } &:&=e_{\mu }^{\text{ }i}s_{M\Sigma \ast }\partial _{i}=\partial
_{\xi ^{\mu }}-e_{\mu }^{\text{ }i}\widetilde{e}_{i}\text{,} \\
&&  \notag \\
h_{\mu } &:&=h_{\mu }^{\text{ }i}s_{M\Sigma \ast }\partial _{i}=\partial
_{\zeta ^{\mu }}-h_{\mu }^{\text{ }i}\widetilde{h}_{i}\text{,}
\end{eqnarray}%
with corresponding tetrad-like components%
\begin{eqnarray}
e_{i}^{\text{ }\mu }\left( \xi \right) &=&e^{\phi }\left( \upsilon _{i}^{%
\text{ }\mu }\left( \xi \right) +r_{\text{ \ }\sigma }^{\alpha }\overset{%
\text{C}}{\Gamma }\text{ }_{i}^{\text{\ }\sigma }\mathfrak{B}_{\alpha }^{%
\text{ }\mu }\left( \xi \right) \right) \text{,}  \label{tetrad} \\
&&  \notag \\
h_{i}^{\text{ }\mu }\left( \xi \text{, }\zeta \right) &=&e^{-\phi }\left(
\upsilon _{\rho }^{\mu }\left( \zeta \right) +\upsilon _{i}^{\sigma }\left(
\xi \right) \mathfrak{B}_{\sigma }^{\text{ \ }\mu }\left( \zeta \right)
\right) \text{,}
\end{eqnarray}%
and basis vectors (on $M$)%
\begin{equation}
\widetilde{e}_{j}\left( \xi \right) =\widetilde{c}_{M\Sigma \ast }\partial
_{j}-e^{\phi }\left[ r_{\mu }^{\text{ \ }\nu }\left( \overset{\text{GL}}{%
\Gamma }\text{ }_{j\alpha }^{\text{ \ \ \ }\mu }\xi ^{\alpha }+\overset{%
\text{D}}{\Gamma }_{j}\xi ^{\mu }+\overset{\text{T}}{\Gamma }\text{ }_{j}^{%
\text{ }\mu }\right) +\overset{\text{C}}{\Gamma }\text{ }_{j}^{\text{\ }%
\sigma }r_{\text{ \ }\sigma }^{\mu }\mathfrak{B}_{\mu }^{\text{ }\nu }\left(
\xi \right) \right] \partial _{\xi ^{\nu }}
\end{equation}%
and%
\begin{equation}
\widetilde{h}_{j}\left( \xi \text{, }\zeta \right) =\widetilde{c}_{M\Sigma
\mathbb{\ast }}\partial _{j}+e^{-\phi }\left[ r_{\text{ \ }\rho }^{\mu
}\left( \overset{\text{GL}}{\Gamma }\text{ }_{j\alpha }^{\text{\ \ \ \ }\rho
}\zeta ^{\alpha }+\overset{\text{C}}{\Gamma }\text{ }_{j}^{\text{ }\rho
}\right) +r_{\text{ \ }\sigma }^{\gamma }\left( \overset{\text{GL}}{\Gamma }%
\text{ }_{j\alpha }^{\text{ \ \ \ }\sigma }\xi ^{\alpha }+\overset{\text{D}}{%
\Gamma }_{j}\xi ^{\sigma }+\overset{\text{T}}{\Gamma }\text{ }_{j}^{\text{ }%
\sigma }\right) \mathfrak{B}_{\gamma }^{\mu }\left( \zeta \right) \right]
\partial _{\zeta ^{\mu }}\text{.}
\end{equation}%
Here $\upsilon ^{\beta }\left( \zeta \right) =\upsilon ^{\beta }\left( \xi
\rightarrow \zeta \right) $, \ $\mathfrak{B}_{\alpha }^{\beta }\left( \zeta
\right) =\mathfrak{B}_{\alpha }^{\rho }\left( \xi \rightarrow \zeta \right) $%
. By definition, the basis vectors satisfy the orthogonality relations%
\begin{equation}
\left\langle V_{\Sigma }^{\mu }|\widetilde{e}_{j}\right\rangle =0\text{,\ }%
\left\langle \vartheta _{\Sigma }^{\mu }|\widetilde{h}_{j}\right\rangle =0%
\text{, }\left\langle V^{\mu }|e_{\nu }\right\rangle =\delta _{\nu }^{\mu }%
\text{,\ }\left\langle \vartheta ^{\mu }|h_{\nu }\right\rangle =\delta _{\nu
}^{\mu }\text{.}
\end{equation}%
We introduce the dilatonic and symmetric $GL_{4}$ \textit{algebra} bases%
\begin{equation}
\flat :=\partial _{\phi }-d^{i}\widetilde{d}_{i}\text{,}\ \ f_{\mu \nu
}:=\partial _{\alpha ^{\mu \nu }}-f_{\mu \nu }^{\text{ }i}\widetilde{f}_{i}
\end{equation}%
with \textit{auxiliary} \textit{soldering} components $d_{i}$ and $f_{i}^{%
\text{ }\mu \nu }$,%
\begin{eqnarray}
d_{i} &=&\zeta _{\sigma }r_{\text{ \ }\rho }^{\sigma }\left( \overset{\text{%
GL}}{^{\dagger }D_{i}}\xi ^{\rho }+\overset{\text{D}}{\Gamma }_{i}\xi ^{\rho
}+\overset{\text{T}}{\Gamma }\text{ }_{i}^{\rho }+\overset{\text{C}}{\Gamma }%
\text{ }_{i}^{\rho }\right) -\frac{1}{2}\partial _{i}\phi \text{,} \\
&&  \notag \\
f_{i}^{\text{ }\mu \nu } &=&\left( r^{-1}\right) _{\;\sigma }^{\mu }\overset{%
\text{GL}}{\Gamma }\text{ }_{i}^{\sigma \beta }r_{\beta }^{\;\nu }-\left(
r^{-1}\right) _{\;\sigma }^{\mu }\partial _{i}r^{\sigma \nu }\text{.}
\end{eqnarray}%
The \textit{coordinate} bases $\widetilde{d}_{j}$ and $\widetilde{f}_{j}$
read%
\begin{equation}
\widetilde{d}_{j}\left( \xi \text{, }\zeta \text{, }\phi \text{, }h\right) :=%
\widetilde{c}_{M\Sigma \ast }\partial _{j}-\zeta _{\sigma }r_{\text{ \ }\rho
}^{\sigma }\left( \overset{\text{GL}}{^{\dagger }\Gamma }\text{ }_{\text{ }%
j\gamma }^{\rho }\xi ^{\gamma }+\overset{\text{D}}{\Gamma }_{j}\xi ^{\rho }+%
\overset{\text{T}}{\Gamma }\text{ }_{j}^{\rho }+\overset{\text{C}}{\Gamma }%
\text{ }_{j}^{\rho }\right) \partial _{\phi }\text{,}
\end{equation}%
and%
\begin{equation}
\widetilde{f}_{j}\left( \xi \text{, }h\right) :=\widetilde{c}_{M\Sigma \ast
}\partial _{j}-\left( \left( r^{-1}\right) _{\;\ \ \ \sigma }^{(\mu |}%
\overset{\text{GL}}{\Gamma }\text{ }_{j}^{\text{ \ }\sigma \beta }r_{\beta
}^{\;\ |\nu )}-\left( r^{-1}\right) _{\;\ \ \ \sigma }^{(\mu |}\partial
_{j}r^{\sigma |\nu )}\right) \partial _{h^{\mu \nu }}\text{.}
\end{equation}%
The bases satisfy%
\begin{equation}
\left\langle \Phi |\widetilde{d}_{i}\right\rangle =0\text{, }\left\langle
\Upsilon ^{\alpha \beta }|\widetilde{f}_{i}\right\rangle =0\text{, }%
\left\langle \Phi |\flat \right\rangle =I\text{,\ }\left\langle \Upsilon
^{\alpha \beta }|f_{\mu \nu }\right\rangle =\delta _{\mu }^{\alpha }\delta
_{\nu }^{\beta }\text{.}
\end{equation}%
With the basis vectors and tetrad components in hand, we observe%
\begin{equation}
\begin{array}{c}
V_{M}^{\mu }:=dx^{i}\otimes e_{i}^{\mu }\text{,}\ \vartheta _{M}^{\mu
}:=dx^{i}\otimes h_{i}^{\mu }\text{,} \\
\\
\Phi _{M}:=dx^{i}\otimes e_{i}^{\alpha }\left\langle \Phi |e_{\alpha
}\right\rangle =dx^{i}\otimes d_{i}\text{.}%
\end{array}
\label{inter4}
\end{equation}%
The symmetric and antisymmetric $GL_{4}$ connection pulled back to $M$ is
given by%
\begin{equation}
\left.
\begin{array}{c}
\Upsilon _{M}^{\mu \nu }=dx^{i}\otimes e_{i}^{\alpha }\left\langle \Upsilon
_{\Sigma }^{\mu \nu }|e_{\alpha }\right\rangle :=dx^{i}\otimes f_{i}^{\text{
}\mu \nu }\text{,} \\
\\
\overset{\circ }{\Gamma }\text{ }_{M}^{\mu \nu }=dx^{i}\otimes e_{i}^{\alpha
}\left\langle \overset{\circ }{\Gamma }\text{ }_{\Sigma }^{\mu \nu
}|e_{\alpha }\right\rangle :=dx^{i}\otimes \overset{\circ }{\Gamma }\text{ }%
_{i}^{\mu \nu }\text{.}%
\end{array}%
\right.  \label{inter5}
\end{equation}%
With the aid of (\ref{inter4}) and (\ref{inter5}), we determine%
\begin{equation}
V_{i}^{\beta }:=e_{i}^{\text{ }\alpha }\left\langle V_{\Sigma }^{\beta
}|e_{\alpha }\right\rangle =e_{i}^{\text{ }\alpha }\delta _{\alpha }^{\beta
}=e_{i}^{\text{ }\beta }\text{, }\vartheta _{i}^{\beta }\equiv h_{i}^{\beta }%
\text{, }\Upsilon _{i}^{\mu \nu }\equiv f_{i}^{\text{ }\mu \nu }\text{, }%
\Phi _{i}\equiv d_{i}\text{.}
\end{equation}

The horizontal tangent subspace vectors in $\widetilde{\pi }_{\mathbb{P}%
\Sigma }:\mathbb{P\rightarrow }$ $\Sigma $ are given by%
\begin{equation}
\widehat{E}_{i}=\widetilde{c}_{M\mathbb{P\ast }}\widetilde{e}_{i}+i%
\widetilde{c}_{M\Sigma \mathbb{\ast }}\left\langle \overset{\circ }{\Gamma }%
\text{ }^{\alpha \beta }|\widetilde{e}_{i}\right\rangle \overset{\text{Int}}{%
\widehat{\mathfrak{R}}\text{ }_{\alpha \beta }^{\left( \mathbf{L}\right) }}%
\text{,}  \label{hor1}
\end{equation}%
\begin{equation}
\widehat{E}_{\mu }=\widetilde{c}_{\Sigma \mathbb{P\ast }}\widetilde{e}_{\mu
}+i\left\langle \overset{\circ }{\Gamma }\text{ }^{\alpha \beta }|\widetilde{%
e}_{\mu }\right\rangle \overset{\text{Int}}{\widehat{\mathfrak{R}}\text{ }%
_{\alpha \beta }^{\left( \mathbf{L}\right) }}\text{,}  \label{hor2}
\end{equation}%
and satisfy%
\begin{equation}
\left\langle \overset{\mathbf{L}}{\omega }|\widehat{E}_{j}\right\rangle
=0=\left\langle \overset{\mathbf{L}}{\omega }|\widehat{E}_{\mu
}\right\rangle \text{.}
\end{equation}%
The right invariant fundamental vector operator\emph{\ }appearing in (\ref%
{hor1})\emph{\ }or (\ref{hor2}) is given by%
\begin{equation}
\widehat{\mathfrak{R}}\text{ }_{\mu \nu }^{\left( \mathbf{L}\right)
}=i\left( \widetilde{\beta }_{[\mu |}^{\text{ \ \ \ \ }\gamma }\frac{%
\partial }{\partial \widetilde{\beta }^{|\nu ]\gamma }}+\epsilon _{\lbrack
\mu }\frac{\partial }{\partial \epsilon ^{\nu ]}}\right) \text{.}
\end{equation}%
On the other hand, the vertical tangent subspace vector in $\widetilde{\pi }%
_{\mathbb{P}\Sigma }:\mathbb{P\rightarrow }$ $\Sigma $ satisfies%
\begin{equation}
\left\langle \overset{\mathbf{L}}{\omega }|\widehat{\mathfrak{L}}\text{ }%
_{\mu \nu }^{\left( \mathbf{L}\right) }\right\rangle =\mathbf{L}_{\mu \nu
}=\left\langle \overset{\mathbf{L}}{\omega }|\widehat{\mathfrak{R}}\text{ }%
_{\mu \nu }^{\left( \mathbf{L}\right) }\right\rangle \text{,}
\end{equation}%
where%
\begin{equation}
\widehat{\mathfrak{L}}\text{ }_{\mu \nu }^{\left( \mathbf{L}\right) }=i%
\widetilde{\beta }_{\gamma \lbrack \mu |}\frac{\partial }{\partial
\widetilde{\beta }_{\gamma }^{\text{ \ }|\nu ]}}\text{, }\widehat{\mathfrak{R%
}}\text{ }_{\mu \nu }^{\left( \mathbf{L}\right) }=i\left( \widetilde{\beta }%
_{[\mu |}^{\text{ \ \ \ \ }\gamma }\frac{\partial }{\partial \widetilde{%
\beta }^{|\nu ]\gamma }}+\epsilon _{\lbrack \mu }\frac{\partial }{\partial
\epsilon ^{\nu ]}}\right) \text{.}
\end{equation}%
and $\widetilde{\beta }_{\mu }^{\text{ }\nu }:=e^{\beta _{\mu }^{\text{ }\nu
}}=\delta _{\mu }^{\text{ }\nu }+\beta _{\mu }^{\text{ }\nu }+\frac{1}{2!}%
\beta _{\mu }^{\text{ }\gamma }\beta _{\gamma }^{\text{ }\nu }+\cdot \cdot
\cdot $. The horizontal tangent subspace vectors in $\Pi _{\Sigma M}:\Sigma
\mathbb{\rightarrow }M$ are given by%
\begin{equation}
\widetilde{E}_{j}=\widetilde{c}_{\Sigma \mathbb{P\ast }}\widetilde{e}_{j}%
\text{,}\ \widetilde{H}_{j}=\widetilde{c}_{\Sigma \mathbb{P\ast }}\widetilde{%
h}_{j}\text{, }\widehat{E}\text{ }_{i}^{\left( \mathbf{D}\right) }=%
\widetilde{c}_{\Sigma \mathbb{P\ast }}\widetilde{d}_{j}\text{,}\ \overset{%
\smile }{E}_{j}=\widetilde{c}_{\Sigma \mathbb{P\ast }}\widetilde{f}_{j}\text{%
,}
\end{equation}%
and satisfy%
\begin{equation}
\left\langle \overset{\mathbf{P}}{\omega }|\widetilde{E}_{j}\right\rangle =0%
\text{, }\left\langle \overset{\mathbf{\Delta }}{\omega }|\widetilde{H}%
_{j}\right\rangle =0\text{, }\left\langle \overset{\text{SY}}{\omega }|%
\overset{\smile }{E}_{j}\right\rangle =0\text{\textbf{,\ }}\left\langle
\overset{\mathbf{D}}{\omega }|\widehat{E}\text{ }_{i}^{\left( \mathbf{D}%
\right) }\right\rangle =0\text{.}
\end{equation}%
The vertical tangent subspace vectors in $\Pi _{\Sigma M}:\Sigma \mathbb{%
\rightarrow }M$ are given by
\begin{equation}
\widetilde{E}_{\mu }=\widetilde{c}_{\Sigma \mathbb{P\ast }}\widehat{%
\mathfrak{L}}\text{ }_{\mu }^{\left( \mathbf{P}\right) }\text{,}\ \overset{%
\smile }{E}_{\alpha \beta }=\widetilde{c}_{\Sigma \mathbb{P\ast }}\widehat{%
\mathfrak{L}}\text{ }_{\alpha \beta }^{\left( \text{SY}\right) }\text{, }%
\widetilde{H}_{\mu }=\widetilde{c}_{\Sigma \mathbb{P\ast }}\widehat{%
\mathfrak{L}}\text{ }_{\mu }^{\left( \mathbf{\Delta }\right) }\text{,}\
\widehat{E}\text{ }^{\left( \mathbf{D}\right) }=\widetilde{c}_{\Sigma
\mathbb{P\ast }}\widehat{\mathfrak{L}}\text{ }^{\left( \mathbf{D}\right) }%
\text{,}  \label{vert}
\end{equation}%
and satisfy%
\begin{equation}
\left\langle \overset{\mathbf{P}}{\omega }|\widetilde{E}_{\mu }\right\rangle
=\mathbf{P}_{\mu }\text{, }\left\langle \overset{\mathbf{\Delta }}{\omega }|%
\widetilde{H}_{\mu }\right\rangle =\mathbf{\Delta }_{\mu }\text{, }%
\left\langle \overset{\text{SY}}{\omega }|\overset{\smile }{E}_{\alpha \beta
}\right\rangle =\text{ }^{\dagger }\mathbf{S}_{\alpha \beta }\text{,\ }%
\left\langle \overset{\mathbf{D}}{\omega }|\widehat{E}\text{ }^{\left(
\mathbf{D}\right) }\right\rangle =\mathbf{D}\text{.}
\end{equation}%
The left invariant fundamental vector operators\emph{\ }appearing in (\ref%
{vert})\emph{\ }are readily computed, the result being%
\begin{equation}
\begin{array}{c}
\widehat{\mathfrak{L}}\text{ }_{\mu }^{\left( \mathbf{P}\right) }=i%
\widetilde{Q}_{\text{ }\mu }^{\nu }\frac{\partial }{\partial \epsilon ^{\nu }%
}\text{, }\widehat{\mathfrak{L}}\text{ }_{\mu }^{\left( \mathbf{\Delta }%
\right) }=i\widetilde{W}_{\text{ }\mu }^{\nu }\frac{\partial }{\partial
b^{\nu }}\text{,} \\
\\
\widehat{\mathfrak{L}}\text{ }_{\alpha \beta }^{\left( \text{SY}\right) }=i%
\widetilde{\alpha }_{\gamma (\mu |}\frac{\partial }{\partial \widetilde{%
\alpha }_{\gamma }^{\text{ \ }|\nu )}}\text{, }\widehat{\mathfrak{L}}\text{ }%
^{\left( \mathbf{D}\right) }=-i\epsilon ^{\beta }\frac{\partial }{\partial
\epsilon ^{\beta }}\text{,}%
\end{array}%
\end{equation}%
where $\widetilde{\alpha }_{\mu }^{\text{ }\nu }:=e^{\alpha _{\mu }^{\text{ }%
\nu }}=\alpha _{\mu }^{\text{ }\nu }+\alpha _{\mu }^{\text{ }\nu }+\frac{1}{%
2!}\alpha _{\mu }^{\text{ }\gamma }\alpha _{\gamma }^{\text{ }\nu }+\cdot
\cdot \cdot $, $\widetilde{Q}_{\sigma }^{\text{ }\alpha }:=\left( \widetilde{%
\chi }_{\sigma }^{\text{ }\alpha }+\delta _{\sigma }^{\text{ }\alpha
}e^{\varphi }\right) $, $\widetilde{W}_{\sigma }^{\text{ }\alpha }:=\left(
\widetilde{\chi }_{\sigma }^{\text{ }\alpha }+\delta _{\sigma }^{\text{ }%
\alpha }e^{-\varphi }\right) $ satisfying $\left( \widetilde{Q}^{-1}\right)
_{\sigma }^{\text{ }\alpha }=\widetilde{Q}_{\text{ }\sigma }^{\alpha }$ and $%
\left( \widetilde{W}^{-1}\right) _{\sigma }^{\text{ }\alpha }=\widetilde{W}_{%
\text{ }\sigma }^{\alpha }$. Making use of the transformation law of the
nonlinear connection (\ref{NLR-transf}) we obtain%
\begin{equation}
\delta \Gamma =\delta V^{\alpha }\mathbf{P}_{\alpha }+\delta \vartheta
^{\alpha }\mathbf{\Delta }_{\alpha }+2\delta \Phi \mathbf{D}+\delta \mathbf{%
\Gamma }^{\alpha \beta }\text{ }^{\dagger }\mathbf{\Lambda }_{\alpha \beta }
\end{equation}%
where%
\begin{equation}
\delta V^{\nu }=u_{\alpha }^{\text{ \ }\nu }V^{\alpha }\text{,}\ \delta
\vartheta ^{\nu }=u_{\alpha }^{\text{ \ }\nu }\vartheta ^{\alpha }\text{, }%
\delta \Phi =0\text{,}\ \delta \mathbf{\Gamma }^{\alpha \beta }=\text{ }%
^{\dagger }\overset{\text{GL}}{\nabla }u^{\alpha \beta }\text{.}
\label{conn-var}
\end{equation}%
From $\delta \mathbf{\Gamma }^{\alpha \beta }=$ $^{\dagger }\overset{\text{GL%
}}{\nabla }u^{\alpha \beta }$ we observe that%
\begin{equation}
\delta \Gamma ^{\lbrack \alpha \beta ]}=\overset{\circ }{\nabla }u^{\alpha
\beta }\text{, }\delta \Upsilon _{\alpha \beta }=2u^{\rho }{}_{(\alpha
|}\Upsilon _{\rho |\beta )}\text{.}  \label{nl-pot}
\end{equation}%
According to (\ref{conn-var}), the nonlinear translational and special
conformal gauge fields transform as contravariant vector valued 1-forms
under $H$, the antisymmetric part of $\mathbf{\Gamma }^{\alpha \beta }$\
transforms inhomogeneously as a gauge potential and the nonlinear dilaton
gauge field $\Phi $ transforms as a scalar valued 1-form. From (\ref{nl-pot}%
) it is clear that the symmetric part of $\mathbf{\Gamma }^{\alpha \beta }$\
is a tensor valued 1-form. Being $4$-covectors we identify $V^{\nu }$ as
coframe fields. The connection coefficient $\overset{\circ }{\Gamma }$ $%
^{\alpha \beta }$ serves as the gravitational gauge potential. The
remaining components of $\mathbf{\Gamma }$, namely $\vartheta $,
$\Upsilon $ and $\Phi $ are dynamical fields of the theory. As
will be seen in the following Section, the tetrad components of
the coframe are used in conjunction with the $H$-metric to induce
a spacetime metric on $M$.

At this point, we have discussed all the mathematical tools that
we will use to realize the Invariance Induced Gravity. In next
Sections, we will proceed with the program of constructing the
induced metric, the action functional and the field equations.
This will be the original contribution of the present review
article where we intend to give a comprehensive approach to
gravity derived from group deformations and conformal-affine
transformations.

\section{The Induced Metric}

The bundle structure of gravitation, together with the
conformal-affine algebra and the nonlinear realizations of gauge
transformations (in particular the classification of  gauge
potentials), provide us all the tools to realize the Invariance
Induced Gravity. In the following part of the paper, we will
derive  the gravitational field and internal symmetry (spin)
quantities showing that they are nothing else but realizations of
the local conformal-affine transformations. In other words, the
deformations of the Poincar\'{e} group give rise to gravity and
internal symmetries (see also \cite{felix,deformazioni}.

Since the Lorentz group $H$ is a subgroup of $G$, we inherit the invariant ($%
\delta o_{\alpha \beta }=\delta o^{\alpha \beta }=0$) (constant) metric of $%
H $, where $o^{\alpha \beta }=o_{\alpha \beta }=diag\left( -\text{, }+\text{%
, }+\text{, }+\right) $. With the aid of $o_{\alpha \beta }$ and the tetrad
components $e_{i}^{\text{ }\alpha }$ given in (\ref{tetrad}), we define the
spacetime metric
\begin{equation}
g_{ij}=e_{i}^{\text{ }\alpha }e_{j}^{\text{ }\beta }o_{\alpha \beta }\text{.}
\end{equation}%
Observing $\overset{\text{GL}}{^{\dagger }\nabla }o_{\alpha \beta
}=-2\Upsilon _{\alpha \beta }$ (where we used $do_{\alpha \beta }=0$) and
taking account of the (second) transformation property (\ref{nl-pot}), we
interpret $\Upsilon _{\alpha \beta }$ as a sort of nonmetricity, i.e. a
deformation (or distortion) gauge field that describes the difference
between the general linear connection and the Levi-Civita connection of
Riemannian geometry \cite{deformazioni}.
In the limit of vanishing gravitational interactions, we have $%
\overset{\text{T}}{\Gamma }$ $^{\sigma }\sim \overset{\text{C}}{\Gamma }$ $%
^{\sigma }\sim \overset{\circ }{\Gamma }$ $_{\text{ }\beta }^{\alpha }\sim
\Upsilon _{\text{ }\beta }^{\alpha }\sim \Phi \rightarrow 0$, $r_{\sigma
}^{\beta }\rightarrow \delta _{\sigma }^{\beta }$ (to first order) and $%
\overset{\text{GL}}{^{\dagger }D}\xi ^{\sigma }\rightarrow d\xi ^{\sigma }$.
Under these conditions, the coframe reduces to $V^{\beta }\rightarrow
e^{\phi }\delta _{\alpha }^{\beta }d\xi ^{\alpha }$ leading to the spacetime
metric%
\begin{equation}
g_{ij}\rightarrow e^{2\phi }\delta _{\alpha }^{\rho }\delta _{\beta
}^{\sigma }\left( \partial _{i}\xi ^{\alpha }\right) \left( \partial _{j}\xi
^{\beta }\right) o_{\rho \sigma }=e^{2\phi }\left( \partial _{i}\xi ^{\alpha
}\right) \left( \partial _{j}\xi ^{\beta }\right) o_{\alpha \beta }
\end{equation}%
characteristic of a Weyl geometry. In this sense, the invariance
properties induce the gravitational field  and generalize results
in \cite{deformazioni}. It is worth noting that conformal
transformations of the metric tensor constitute only a part of the
whole deformation field.

\section{The Cartan Structure Equations}
Our task is now to deduce the dynamics. Using the nonlinear gauge
potentials derived in  Eqs.(\ref{NL-p2}), (\ref{NL-p3}),
(\ref{NL-p4}), the covariant derivative defined on $\Sigma $
pulled back to $M$ has the form
\begin{equation}
\mathbf{\nabla }:=d-iV^{\alpha }\mathbf{P}_{\alpha }-i\vartheta ^{\alpha }%
\mathbf{\Delta }_{\alpha }-2i\Phi \mathbf{D}-i\Gamma ^{\alpha \beta }\text{ }%
^{\dagger }\mathbf{\Lambda }_{\alpha \beta }.  \label{NL-covDer}
\end{equation}%
By using  (\ref{NL-covDer}) together with the relevant Lie algebra
commutators, we obtain the the bundle curvature%
\begin{equation}
\mathbb{F}:=\mathbf{\nabla }\wedge \mathbf{\nabla }=-i\mathcal{T}^{\alpha }%
\mathbf{P}_{\alpha }-i\mathcal{K}^{\alpha }\mathbf{\Delta }_{\alpha }-i%
\mathcal{Z}\mathbf{D}-i\mathbb{R}_{\alpha }^{\text{ \ }\beta }\text{ }%
^{\dagger }\mathbf{\Lambda }_{\text{ \ }\beta }^{\alpha }\text{.}
\end{equation}%
The field strength components of $\mathbb{F}$\ are given by the first Cartan
structure equations. They are respectively, the projectively deformed, $%
\Upsilon $-distorted translational field strength%
\begin{equation}
\mathcal{T}^{\alpha }:=\text{ }^{\dagger }\overset{\text{GL}}{\nabla }%
V^{\alpha }+2\Phi \wedge V^{\alpha },
\end{equation}%
the projectively deformed, $\Upsilon $-distorted special conformal field
strength%
\begin{equation}
\mathcal{K}^{\alpha }:=\text{ }^{\dagger }\overset{\text{GL}}{\nabla }%
\vartheta ^{\alpha }-2\Phi \wedge \vartheta ^{\alpha },
\end{equation}%
the $\Psi $-deformed Weyl homothetic curvature 2-form (dilaton field
strength)%
\begin{equation}
\mathcal{Z}:=d\Phi +\Psi \text{,}\ \Psi =V\cdot \vartheta -\vartheta \cdot V
\end{equation}%
and the general conformal-affine curvature%
\begin{equation}
\mathbb{R}^{\alpha \beta }:=\widehat{R}\text{ }^{\alpha \beta }+\Psi
^{\alpha \beta }\text{,}
\end{equation}%
with%
\begin{equation}
\widehat{R}\text{ }^{\alpha \beta }:=\mathfrak{R}^{\alpha \beta }+\mathcal{R}%
^{\alpha \beta }\text{, \ }\Psi ^{\alpha \beta }:=V^{[\alpha }\wedge
\vartheta ^{\beta ]}\text{.}  \label{affine-curv}
\end{equation}%
Operator $^{\dagger }\overset{\text{GL}}{\nabla }$ denotes the nonlinear
covariant derivative built from volume preserving (VP) connection (i.e.
excluding $\Phi $) forms respectively. The $\Upsilon $ and $\overset{\circ }{%
\Gamma }$-affine curvatures in (\ref{affine-curv}) read%
\begin{eqnarray}
\mathfrak{R}^{\alpha \beta } &:&=\overset{\circ }{\nabla }\Upsilon ^{\alpha
\beta }+\Upsilon _{\gamma }^{\alpha }\wedge \Upsilon ^{\gamma \beta }\text{,}
\\
&&  \notag \\
\mathcal{R}^{\alpha \beta } &:&=d\overset{\circ }{\Gamma }\text{ }^{\alpha
\beta }+\overset{\circ }{\Gamma }\text{ }_{\gamma }^{\text{ }\alpha }\wedge
\overset{\circ }{\Gamma }\text{ }^{\gamma \beta }\text{,}
\end{eqnarray}%
respectively. Operator $\overset{\circ }{\nabla }$ is defined with respect
to the restricted connection $\overset{\circ }{\Gamma }$ $^{\alpha \beta }$
given in (\ref{rest}).

The field strength components of the bundle curvature have the following
group variations%
\begin{equation}
\delta \mathbb{R}_{\alpha }^{\text{ }\beta }=u_{\alpha }^{\text{ }\gamma }%
\mathbb{R}_{\text{ \ }\gamma }^{\beta }-u_{\gamma }^{\text{ }\beta }\mathbb{R%
}_{\alpha }^{\text{ }\gamma }\text{, }\delta \mathcal{Z}=0\text{, }\delta
\mathcal{T}^{\alpha }=-u_{\beta }^{\text{ }\alpha }\mathcal{T}^{\beta }\text{%
,\ }\delta \mathcal{K}^{\alpha }=-u_{\beta }^{\text{ }\alpha }\mathcal{K}%
^{\beta }\text{.}
\end{equation}%
A gauge field Lagrangian is built from polynomial combinations of the
strength $\mathbb{F}$ defined as
\begin{equation}
\mathbb{F}\left( \Gamma \left( \Omega \text{, }D\xi \right) \text{, }d\Gamma
\right) :=\nabla \wedge \nabla =d\Gamma +\Gamma \wedge \Gamma \text{.}
\end{equation}
Now we have all the ingredients to derive the conservation laws
that constitute a fundamental result of our approach rendering the
theory self-consistent.

\section{The Bianchi Identities}

In what follows, the Bianchi identities  play a central role being
the conservation laws of the theory. We therefore derive them
presently.

1a) The $1^{st}$ translational Bianchi identity reads,%
\begin{equation}
\overset{\text{GL}}{\nabla }\mathcal{T}^{a}=\widehat{R}\text{ }_{\text{ }%
\beta }^{\alpha }\wedge V^{\beta }+\Phi \wedge T^{a}+2d\left( \Phi \wedge
V^{\alpha }\right) \text{.}
\end{equation}

1b) Similarly to the case in (1a), the $1^{st}$ conformal Bianchi
identities are
respectively given by%
\begin{equation}
\overset{\text{GL}}{\nabla }\mathcal{K}^{a}=\widehat{R}\text{ }_{\text{ }%
\beta }^{\alpha }\wedge \vartheta ^{\beta }-\Phi \wedge \mathcal{K}%
^{a}-2d\left( \Phi \wedge \vartheta ^{\alpha }\right) \text{.}
\end{equation}%
2a) The $\Upsilon $ and $\overset{\circ }{\Gamma }$-affine component of the $%
2^{nd}$ Bianchi identity is given by%
\begin{equation}
^{\dagger }\overset{\text{GL}}{\nabla }\mathfrak{R}^{\alpha \beta }=2%
\mathfrak{R}_{\text{ \ \ \ }\gamma }^{(\alpha |}\Upsilon ^{\gamma |\beta )}%
\text{, }^{\dagger }\overset{\text{GL}}{\nabla }\mathcal{R}^{\alpha \beta }=0%
\text{,}
\end{equation}%
respectively. Hence, the generalized $2^{nd}$ Bianchi identity is given by%
\begin{equation}
^{\dagger }\overset{\text{GL}}{\nabla }\widehat{R}\text{ }_{\text{ }\beta
}^{\alpha }=2\mathfrak{R}_{\text{ \ \ \ \ }\gamma }^{(\alpha |}\Upsilon
^{\gamma |\rho )}o_{\rho \beta }\text{.}
\end{equation}%
Since the full curvature $\mathbb{R}^{\alpha \beta }$ is proportional to $%
\Psi ^{\alpha \beta }$, it is necessary to consider%
\begin{equation}
^{\dagger }\overset{\text{GL}}{\nabla }\Psi ^{\alpha \beta }=\text{ }%
^{\dagger }\mathcal{T}^{\alpha }\wedge \vartheta ^{\beta }+V^{\alpha }\wedge
\text{ }^{\dagger }\mathcal{K}^{\beta }\text{,}
\end{equation}%
from which we conclude%
\begin{equation}
^{\dagger }\overset{\text{GL}}{\nabla }\mathbb{R}^{\alpha \beta }=2\mathfrak{%
R}_{\text{ \ \ \ }\gamma }^{(\alpha |}\Upsilon ^{\gamma |\beta )}+\text{ }%
^{\dagger }\mathcal{T}^{\alpha }\wedge \vartheta ^{\beta }+V^{\alpha }\wedge
\text{ }^{\dagger }\mathcal{K}^{\beta }.
\end{equation}%
2c) The dilatonic component of the $2^{nd}$ Bianchi identity is given by%
\begin{equation}
\overset{\text{GL}}{\nabla }\mathcal{Z}=dZ+\overset{\text{GL}}{\nabla }%
\left( V\wedge \vartheta \right) =\overset{\text{GL}}{\nabla }\Psi +\Phi
\wedge \Psi \text{,}
\end{equation}%
From the definition of $\Psi $, we obtain%
\begin{equation}
\nabla \Psi =\mathcal{T}^{\alpha }\wedge \vartheta _{\alpha }+V_{\alpha
}\wedge \mathcal{K}^{\alpha }+\Phi \wedge \left( V_{\alpha }\wedge \vartheta
^{\alpha }\right) \text{.}
\end{equation}%
Defining%
\begin{equation}
\Sigma ^{\mu \nu }:=\mathbf{B}^{\mu \nu }+\Psi ^{\mu \nu }\text{,}\ \mathbf{B%
}^{\mu \nu }:=B^{\mu \nu }+\mathcal{B}^{\mu \nu }\text{, }B^{\mu \nu
}:=V^{\mu }\wedge V^{\nu }\text{, \ }\mathcal{B}^{\mu \nu }:=\vartheta ^{\mu
}\wedge \vartheta ^{\nu }\text{,}
\end{equation}%
and asserting $V^{\alpha }\wedge \vartheta _{\alpha }=0$, we find $\Sigma
_{\mu \nu }\wedge \Sigma ^{\mu \nu }=0$. Using this result,we obtain%
\begin{equation}
\nabla \Psi =\mathcal{T}^{\alpha }\wedge \vartheta _{\alpha }+V_{\alpha
}\wedge \mathcal{K}^{\alpha }\text{.}
\end{equation}
The last step is now to derive the field equations.

\section{The Action Functional and the Field Equations}

We seek an action for a local gauge theory based on the $CA\left( 3\text{, }%
1\right) $ symmetry group. We consider the $3D$ topological invariants $%
\mathbb{Y}$ of the non-Riemannian manifold of conformal-affine
connections. Our objective is the $4D$ boundary terms $\mathbb{B}$
obtained by means of exterior differentiation of these $3D$
invariants, i.e. $\mathbb{B}=d\mathbb{Y}$. The Lagrangian density
of conformal-affine gravity is modelled after $\mathbb{B}$, with
appropriate distribution of Lie star operators so as to
re-introduce the
dual frame fields. The generalized conformal-affine surface topological invariant reads%
\begin{equation}
\mathbb{Y}=-\frac{1}{2l^{2}}\left[
\begin{array}{c}
\theta _{\mathcal{A}}\left( \mathcal{A}_{a}^{\text{ }b}\wedge \widehat{R}%
\text{ }_{b}^{\text{ }a}+\frac{1}{3}\mathcal{A}_{a}^{\text{ }b}\wedge
\mathcal{A}_{b}^{\text{ }c}\wedge \mathcal{A}_{c}^{\text{ }a}\right) + \\
\\
-\theta _{\mathcal{V}}\mathcal{V}_{a}\wedge \mathbf{T}^{\alpha }+\theta
_{\Phi }\Phi \wedge \mathcal{Z}%
\end{array}%
\right] \text{,}
\end{equation}%
where $\mathbf{T}^{\alpha }:=\mathcal{T}^{\alpha }+\mathcal{K}^{\alpha }$.
The associated total conformal-affine boundary term  is given by,%
\begin{equation}
\mathbb{B}=\frac{1}{2l^{2}}\left[
\begin{array}{c}
\widehat{R}_{\beta \alpha }\wedge \mathbf{B}^{\beta \alpha }+\Sigma
^{\lbrack \beta \alpha ]}\wedge \Sigma _{\lbrack \beta \alpha ]}-\widehat{R}%
\text{ }^{\alpha \beta }\wedge \widehat{R}_{\alpha \beta }-\mathcal{Z}\wedge
\mathcal{Z}+ \\
\\
+\mathcal{K}_{\alpha }\wedge \mathcal{K}^{\alpha }+\mathcal{T}_{\alpha
}\wedge \mathcal{T}^{\alpha }-\Phi \wedge \left( V_{\alpha }\wedge \mathcal{T%
}^{\alpha }+\vartheta _{\alpha }\wedge \mathcal{K}^{\alpha }\right) + \\
\\
-\Upsilon _{\alpha \beta }\wedge \left( V^{\alpha }\wedge \mathcal{T}^{\beta
}+\vartheta ^{\alpha }\wedge \mathcal{K}^{\beta }\right) \text{.}%
\end{array}%
\right]  \label{boundary}
\end{equation}

Using the boundary term (\ref{boundary}) as a guide, we choose $[48$, $51$, $%
54$, $56,$ $66]$ an action of form%
\begin{equation}
I=\int_{\mathcal{M}}\left\{
\begin{array}{c}
d\left( \mathcal{V}^{\alpha }\wedge \mathbf{T}_{\alpha }\right) +\widehat{R}%
\text{ }^{\alpha \beta }\wedge \Sigma _{\star \alpha \beta }+\mathcal{B}%
_{\star \alpha \beta }\wedge \mathcal{B}^{\alpha \beta }+\Psi _{\star \alpha
\beta }\wedge \Psi ^{\alpha \beta }+\eta _{\star \alpha \beta }\wedge \eta
^{\alpha \beta } \\
\\
-\frac{1}{2}\left( \mathcal{R}_{\star \mu \nu }\wedge \mathcal{R}^{\mu \nu }+%
\mathcal{Z}\wedge \star \mathcal{Z}\right) +\mathcal{T}_{\star \alpha
}\wedge \mathcal{T}^{\alpha }+\mathcal{K}_{\star \alpha }\wedge \mathcal{K}%
^{\alpha }+ \\
\\
-\Phi \wedge \left( \mathcal{T}^{\star \alpha }\wedge V_{\alpha }+\mathcal{K}%
^{\star \alpha }\wedge \vartheta _{\alpha }\right) -\Upsilon _{\alpha \beta
}\wedge \left( V^{\alpha }\wedge \mathcal{T}^{\star \beta }+\vartheta
^{\alpha }\wedge \mathcal{K}^{\star \beta }\right) \text{.}%
\end{array}%
\right\}  \label{action}
\end{equation}%
Note that the action integral (\ref{action}) is invariant under
Lorentz rather than conformal-affine transformations. The Lie star
$\star $ operator is defined as $\star V_{\alpha
}=\frac{1}{3!}\eta _{\alpha \beta \mu \nu }V^{\beta }\wedge V^{\mu
}\wedge V^{\nu }$.

The field equations are obtained from the variation of $I$ with
respect to the independant gauge potentials. It is convenient to
define the functional
derivatives%
\begin{equation}
\begin{array}{c}
\frac{\delta \mathcal{L}_{\text{gauge}}}{\delta V^{\alpha }}:=-\overset{%
\text{GL}}{\nabla }N_{\alpha }+\overset{\text{V}}{\mathfrak{T}}_{\alpha }%
\text{,} \\
\\
\frac{\delta \mathcal{L}_{\text{gauge}}}{\delta \vartheta ^{\alpha }}:=-%
\overset{\text{GL}}{\nabla }M_{\alpha }+\overset{\vartheta }{\mathfrak{T}}%
_{\alpha }\text{,} \\
\\
\mathfrak{Z}_{\alpha }^{\text{ }\beta }:=\frac{\delta \mathcal{L}_{\text{%
gauge}}}{\delta \widehat{\Gamma }\text{ }_{\text{ }\beta }^{\alpha }}=-\text{
}^{\dagger }\overset{\text{GL}}{\nabla }\widehat{M}\text{ }_{\alpha }^{\text{
}\beta }+\widehat{E}\text{ }_{\alpha }^{\text{ }\beta }\text{.}%
\end{array}%
\end{equation}%
where
\begin{equation}
\widehat{M}\text{ }_{\beta }^{\text{ }\alpha }:=-\frac{\partial \mathcal{L}_{%
\text{gauge}}}{\partial \widehat{R}\text{ }_{\text{ }\alpha }^{\beta }}\text{%
, }\widehat{E}\text{ }_{\alpha }^{\text{ }\beta }:=\frac{\partial \mathcal{L}%
_{\text{gauge}}}{\partial \widehat{\Gamma }\text{ }_{\text{ }\beta }^{\alpha
}}\text{, }\overset{\text{V}}{\mathfrak{T}}_{\alpha }:=\frac{\partial
\mathcal{L}_{\text{gauge}}}{\partial V^{\alpha }}\text{, }\overset{\vartheta
}{\mathfrak{T}}_{\alpha }:=\frac{\partial \mathcal{L}_{\text{gauge}}}{%
\partial \vartheta ^{\alpha }}\text{, }\Theta :=\frac{\partial \mathcal{L}_{%
\text{gauge}}}{\partial \Phi }\text{.}
\end{equation}%
The gauge field momenta\textit{\ }are defined by%
\begin{equation}
\begin{array}{c}
N_{\alpha }:=-\frac{\partial \mathcal{L}_{\text{gauge}}}{\partial \mathcal{T}%
^{\alpha }}\text{, }M_{\alpha }:=-\frac{\partial \mathcal{L}_{\text{gauge}}}{%
\partial \mathcal{K}^{\alpha }}\text{, }\Xi :=-\frac{\partial \mathcal{L}_{%
\text{gauge}}}{\partial \mathcal{Z}}\text{,} \\
\\
\widehat{M}_{[\alpha \beta ]}:=N_{\alpha \beta }=-o_{[\alpha |\gamma }\frac{%
\partial \mathcal{L}_{\text{gauge}}}{\partial \mathcal{R}_{\gamma }^{\text{ }%
|\beta ]}}\text{, }\widehat{M}_{(\alpha \beta )}:=M_{\alpha \beta
}=-2o_{(\alpha |\gamma }\frac{\partial \mathcal{L}_{\text{gauge}}}{\partial
\mathfrak{R}_{\gamma }^{\text{ }|\beta )}}\text{.}%
\end{array}%
\end{equation}%
Furthermore, the shear (gauge field deformation)\ and\ hypermomentum
current\ forms are given by%
\begin{equation}
\widehat{E}_{(\alpha \beta )}:=U_{\alpha \beta }=-V_{(\alpha }\wedge \left(
M_{\beta )}+N_{\beta )}\right) -M_{\alpha \beta }\text{, }\widehat{E}%
_{[\alpha \beta ]}:=E_{\alpha \beta }=-V_{[\alpha }\wedge \left( M_{\beta
]}+N_{\beta ]}\right) \text{,}
\end{equation}%
The analogue of the Einstein equations read%
\begin{equation}
G_{\alpha }+\Lambda \widehat{\eta }_{\alpha }+\text{ }^{\dagger }\overset{%
\text{GL}}{\nabla }\mathcal{T}_{\star \alpha }+\overset{\text{V}}{\mathfrak{T%
}}_{\alpha }=0\text{,}
\end{equation}%
with Einstein-like three-form%
\begin{equation}
G_{\alpha }=\left( \mathcal{R}^{\beta \gamma }+\Upsilon _{\text{ \ \ \ }\rho
}^{[\beta |}\wedge \Upsilon ^{|\gamma ]\rho }\right) \wedge \left( \eta
_{\beta \gamma \alpha }+\star \left[ B_{\beta \gamma }\wedge \vartheta
_{\alpha }\right] \right) \text{,}  \label{Einstein}
\end{equation}%
coupling (cosmological) constant $\Lambda $ and mixed three-form
$\widehat{\eta }_{\alpha }=\eta _{\alpha }+\star \left( \vartheta
_{\alpha }\wedge V_{\beta }\right)
\wedge V^{\beta }$. Let us observe that $G_{\alpha }$ includes symmetric $GL_{4}$ $%
\left( \Upsilon \right) $\ as well as special conformal ($\vartheta $)
contributions. The gauge field 3-form $\overset{\text{V}}{\mathfrak{T}}%
_{\alpha }$ is given by%
\begin{eqnarray}
\overset{\text{V}}{\mathfrak{T}}_{\alpha } &=&\left\langle \mathcal{L}_{%
\text{gauge}}|e_{\alpha }\right\rangle +\left\langle \mathcal{Z}|e_{\alpha
}\right\rangle \wedge \Xi +\left\langle \mathcal{T}^{\beta }|e_{\alpha
}\right\rangle \wedge N_{\beta }+  \label{stress} \\
&&  \notag \\
&&+\left\langle \mathcal{K}^{\beta }|e_{\alpha }\right\rangle \wedge
M_{\beta }+\left\langle \mathcal{R}_{\gamma }^{\text{ }\beta }|e_{\alpha
}\right\rangle \wedge N_{\text{ }\beta }^{\gamma }+\frac{1}{2}\left\langle
\mathfrak{R}_{\gamma }^{\text{ }\beta }|e_{\alpha }\right\rangle M_{\text{ }%
\beta }^{\gamma }\text{,}  \notag
\end{eqnarray}%
We remark that to interpret Eqs.(\ref{Einstein}) as the
gravitational field equation analogous to the Einstein equations,
we must transform from the Lie algebra index $\alpha $ to the
spacetime basis index $k$ by contracting over the former $\left(
\alpha \right) $ with the conformal-affine tetrads $e_{k}^{\alpha
}$. This fact is relevant to read gravity in holonomic and
anholonomic frames respectively. It is
\begin{eqnarray}
\overset{\text{V}}{\mathfrak{T}}_{\alpha } &=&\mathfrak{T}_{\alpha }\left[
\mathcal{T}\right] +\mathfrak{T}_{\alpha }\left[ \mathcal{K}\right] +%
\mathfrak{T}_{\alpha }\left[ \mathcal{R}\right] +\mathfrak{T}_{\alpha }\left[
Z\right] -\left\langle \mathcal{T}^{\beta }|e_{\alpha }\right\rangle \wedge
N_{\beta }-\left\langle \mathcal{K}^{\beta }|e_{\alpha }\right\rangle \wedge
M_{\beta }+ \\
&&  \notag \\
&&-\left\langle \mathcal{R}_{\gamma }^{\text{ }\beta }|e_{\alpha
}\right\rangle \wedge N_{\text{ }\beta }^{\gamma }-\left\langle \mathcal{Z}%
|e_{\alpha }\right\rangle \wedge \Xi +\Psi _{\star \alpha \beta }\wedge
\vartheta ^{\beta }+\left\langle \Sigma _{\star \gamma \beta }|e_{\alpha
}\right\rangle \wedge \widehat{R}\text{ }^{\alpha \beta }+  \notag \\
&&  \notag \\
&&+\left\langle \Upsilon ^{\gamma \beta }\wedge \left( V_{\gamma }\wedge
\mathcal{T}_{\star \beta }+\vartheta _{\gamma }\wedge \mathcal{K}_{\star
\beta }\right) |e_{\alpha }\right\rangle +\Sigma _{\star \gamma \beta
}\wedge \left\langle \widehat{R}\text{ }^{\gamma \beta }|e_{\alpha
}\right\rangle +  \notag \\
&&  \notag \\
&&\mathcal{B}_{\star \gamma \beta }\wedge \left\langle \mathcal{B}^{\gamma
\beta }|e_{\alpha }\right\rangle +\left\langle \mathcal{B}_{\star \gamma
\beta }|e_{\alpha }\right\rangle \wedge \mathcal{B}^{\gamma \beta
}+\left\langle \Psi _{\star \gamma \beta }|e_{\alpha }\right\rangle \wedge
\Psi ^{\gamma \beta }  \notag
\end{eqnarray}%
respectively, with
\begin{equation}
\begin{array}{c}
\mathfrak{T}_{\alpha }\left[ \mathcal{R}\right] =\frac{1}{2}a_{1}\left(
\mathcal{R}_{\rho \gamma }\wedge \left\langle \mathcal{R}^{\star \rho \gamma
}|e_{\alpha }\right\rangle -\left\langle \mathcal{R}_{\rho \gamma
}|e_{\alpha }\right\rangle \wedge \mathcal{R}^{\star \rho \gamma }\right)
\text{,} \\
\\
\mathfrak{T}_{\alpha }\left[ \mathcal{T}\right] =\frac{1}{2}a_{2}\left(
\mathcal{T}_{\gamma }\wedge \left\langle \mathcal{T}^{\star \gamma
}|e_{\alpha }\right\rangle -\left\langle \mathcal{T}_{\gamma }|e_{\alpha
}\right\rangle \wedge \mathcal{T}^{\star \gamma }\right) \text{,} \\
\\
\mathfrak{T}_{\alpha }\left[ \mathcal{K}\right] =\frac{1}{2}a_{3}\left(
\mathcal{K}_{\gamma }\wedge \left\langle \mathcal{K}^{\star \gamma
}|e_{\alpha }\right\rangle -\left\langle \mathcal{K}_{\gamma }|e_{\alpha
}\right\rangle \wedge \mathcal{K}^{\star \gamma }\right) \text{,} \\
\\
\mathfrak{T}_{\alpha }\left[ Z\right] =\frac{1}{2}a_{4}\left( d\Phi \wedge
\left\langle \star d\Phi |e_{\alpha }\right\rangle -\left\langle d\Phi
|e_{\alpha }\right\rangle \wedge \star d\Phi \right) \text{.}%
\end{array}%
\end{equation}%
From the variation of $I$ with respect to $\vartheta ^{\alpha }$\, we get%
\begin{equation}
\mathfrak{G}_{\alpha }+\Lambda \widehat{\omega }_{\alpha }+\text{ }^{\dagger
}\overset{\text{GL}}{\nabla }\mathcal{K}_{\star \alpha }+\overset{\vartheta }%
{\mathfrak{T}}_{\alpha }=0\text{,}
\end{equation}%
where, in analogy to Eqs.(\ref{Einstein}), we have%
\begin{equation}
\mathfrak{G}_{\alpha }=h_{i}^{\alpha }\left( \mathcal{R}^{\beta \gamma
}+\Upsilon _{\text{ \ \ \ }\rho }^{[\beta |}\wedge \Upsilon ^{|\gamma ]\rho
}\right) \wedge \left( \omega _{\beta \gamma \alpha }+\star \left[ \mathcal{B%
}_{\beta \gamma }\wedge V_{\alpha }\right] \right) \text{,}
\label{PDEinstein}
\end{equation}%
where $\widehat{\omega }_{\alpha }=\omega _{\alpha }+\star \left( \vartheta
_{\alpha }\wedge V_{\beta }\right) \wedge \vartheta ^{\beta }$. The quantity
$\overset{\vartheta }{\mathfrak{T}}_{i}=h_{i}^{\alpha }\overset{\vartheta }{%
\mathfrak{T}}_{\alpha }$ is similar to (\ref{stress}) but with the algebra
basis $e_{\alpha }$\ replaced by $h_{\alpha }$ and the conformal-affine tetrad components $%
e_{\text{ }i}^{\alpha }$ replaced by $h_{\text{ }i}^{\alpha }$. The two
gravitational field equations (\ref{Einstein}) and (\ref{PDEinstein}) are $%
P-\Delta $ symmetric. We may say that they exhibit $P-\Delta $ duality
symmetry invariance.

From the variational equation for $\overset{\circ }{\Gamma }$ $_{\alpha }^{%
\text{ }\beta }$, we obtain the conformal-affine gravitational
analogue of the
Yang-Mills-torsion type field equation,%
\begin{equation}
\overset{\circ }{\nabla }\star \mathcal{R}_{\alpha }^{\text{ }\beta }+%
\overset{\circ }{\nabla }\star \Sigma _{\alpha }^{\text{ }\beta }+\left(
V^{\beta }\wedge \mathcal{T}_{\star \alpha }+\vartheta ^{\beta }\wedge
\mathcal{K}_{\star \alpha }\right) =0\text{.}  \label{YM}
\end{equation}%
Variation of $I$ with respect to $\Upsilon _{\alpha }^{\text{ }\beta }$
leads to%
\begin{equation}
\overset{\circ }{\nabla }\star \Sigma _{\alpha \beta }-\Upsilon _{(\alpha
|}^{\text{ \ \ \ }\gamma }\wedge \Sigma _{\star \gamma |\beta )}+V_{(\alpha
}\wedge \mathcal{T}_{\star \beta )}+\vartheta _{(\alpha }\wedge \mathcal{K}%
_{\star \beta )}=0\text{.}  \label{shear-eq}
\end{equation}%
Finally, from the variational equation for $\Phi $, the gravi-scalar field
equation is given by%
\begin{equation}
d\star d\Phi +V_{\alpha }\wedge \mathcal{T}^{\star \alpha }+\vartheta
_{\alpha }\wedge \mathcal{K}^{\star \alpha }=0\text{.}  \label{scalar}
\end{equation}

In conclusions, the field equations of conformal-affine gravity
have been obtained in this section. The analogue of the Einstein
equation, obtained from variation of $I$ with respect to the
coframe $V$, is characterized by an Einstein-like 3-form that
includes symmetric $GL_{4}$ as well as special conformal
contributions. Moreover, the field equation in (\ref{Einstein})
contains a non-trivial
torsion contribution. Performing a $P-\Delta $ transformation ( i.e. $%
V\rightarrow \vartheta $, $\mathcal{T}\rightarrow \mathcal{K}$, $%
D\rightarrow -D$) on (\ref{Einstein}) we obtain
(\ref{PDEinstein}). This result may also be obtained directly by
varying $I$ with respect $\vartheta $.
A mixed conformal-affine cosmological constant term arises in (\ref{Einstein}), (\ref{PDEinstein})%
) as a consequence of the structure of the 2-form
$\mathbb{R}_{\text{ }\beta }^{\alpha }$. This result can be
extremely interesting from a physical viewpoint in order to
envisage a mechanism capable of producing the "observed"
cosmological constant (see also \cite{stefano1,stefano2}).

The field equation (\ref{YM}) is a Yang-Mills-like equation that
represents the generalization of the Gauss torsion-free equation
$\nabla \star B^{\alpha \beta }=0$. In our case, we considered a
mixed volume form involving both $V$ and $\vartheta $ leading to
the substitution $B^{\alpha \beta }\rightarrow \Sigma ^{\alpha
\beta }$. Additionally, even in the case of vanishing $T^{\rho
}=\overset{\circ }{\nabla }V^{\rho }$, the conformal-affine
torsion depends on the dilaton potential $\Phi $ which in general
is non-vanishing.
A similar argument holds for the special conformal quantity $\mathcal{K}%
^{\rho }$. Admitting the quadratic curvature term $\mathcal{R}_{\alpha
}^{\beta }\wedge \star \mathcal{R}_{\beta }^{\alpha }$ in the gauge
Lagrangian it becomes clear how we draw the analogy between (\ref{YM}) and
the Gauss equation. Equation (\ref{shear-eq}) follow from similar
considerations as (\ref{YM}), the significant differences being the lack of
a $\overset{\circ }{\nabla }\star \mathfrak{R}_{\alpha }^{\text{ }\beta }$
counterpart to $\overset{\circ }{\nabla }\star \mathcal{R}_{\alpha }^{\text{
}\beta }$ since $\star \mathfrak{R}_{\alpha }^{\text{ }\beta }=0$. Finally, (%
\ref{scalar}) involves both $\mathcal{T}^{\rho }$ and
$\mathcal{K}^{\rho }$ in conjunction with a term that resembles
the source-free Maxwell equations with the dilaton potential
playing a similar role to the electromagnetic vector potential.

\section{Conclusions and Perspectives}

In this review paper, after a summary of the bundle approach to
the gauge theories with a discussion, in particular, of the bundle
structure of gravitation, a nonlinearly realized representation of
the local conformal-affine group has been determined. Before the
physical applications, we have reviewed, in details, all the
mathematical tools to show that gravity and spin are the results
of  the local conformal-affine group so then it is possible to
deal with an Invariance Induced Gravity. It has been found that
the nonlinear Lorentz transformation law contains contributions
from the linear Lorentz parameter as well as conformal and shear
contributions via the nonlinear $4$-boosts and symmetric $GL_{4}$
parameters. We have identified the pullback of the nonlinear
translational connection coefficient to $M$ as a spacetime
coframe. In this way, the frame fields of the theory are obtained
from the (nonlinear) gauge prescription. The mixed index coframe
component (tetrad) is used to convert from Lie algebra indices
into spacetime indices. The spacetime metric is a secondary object
constructed (induced!) from the constant $H$ group metric and the
tetrads. The gauge fields $\overset{\circ }{\Gamma }$ $^{\alpha
\beta }$ are the analogues of the Christoffel connection
coefficients of General Relativity and serve as the gravitational
gauge potentials used to define covariant derivative operators.
The gauge fields $\vartheta $, $\Phi $, and $\Upsilon $ encode
information regarding special conformal, dilatonic and
deformational degrees of freedom of the bundle manifold
\cite{deformazioni}. The spacetime geometry is therefore
determined by gauge field interactions as in  the so called {\it
Emergent Gravity} \cite{seiberg}.

Furthermore, the bundle curvature and the Bianchi identities have
been  determined and then the gauge Lagrangian density have been
modelled after the  boundary topological invariants have been
defined. As a consequence of this approach, no mixed field
strength terms involving different components of the total
curvature arise in the action. The analogue of the Einstein
equations contains a non-trivial torsion contribution which is
directly related to the spin fields of the theory (see also
\cite{cosimo}). The Einstein-like three-form includes symmetric
$GL_{4}$ as well as special conformal contributions. A mixed
translational-conformal cosmological constant term arises due to
the structure of the generalized curvature of the manifold. We
also obtain a Yang-Mills-like equation that represents the
generalization of the Gauss torsion-free equation. Variation of
$I$ with respect to $\Upsilon _{\alpha }^{\text{ }\beta }$ leads
to a constraint equation relating the $GL_{4}$ deformation gauge
field to the translational and special conformal field strengths.
The gravi-scalar field equation has non-vanishing translational
and special conformal contributions. As a concluding remark, we
can say that gravity (and in general any gauge field) can be
derived as the nonlinear realization of a local conformal-affine
symmetry group and then gravity can be considered an interaction
induced from invariance properties. This approach can be adopted
also to generalized theories of gravity
\cite{odintsov,GRG,faraoni} as we are going to do in a forthcoming
paper.

\end{document}